\documentstyle[preprint,prd,aps,floats,rotate,natbib]{revtex}
\tighten			
\def\Ansatz{Ans\"{a}tz}			
\def\Ansatze{Ans\"{a}tze}		
\def\ANSATZE{\text{ANS\"{A}TZE}}
\def\fur{f\"{u}r}
\def\Universitat{Universit\"{a}t}
\def\Gomez{{\text{G\'{o}mez}}}

\def\Hamade{{\text{Hamad\'{e}}}}
\def\Masso{{\text{Mass\'{o}}}}

\def\OMurchadha{{\text{\'{O}~Murchadha}}}

\def\code#1{{\sc #1}}				
\def\defn#1{``#1''}				
\def\xdefn#1{#1}				
\def\nop#1{}					
\def\cf{cf.\hbox{}}
\def\eg{e.g.\hbox{}}

\def\ie{i.e.\hbox{}}

\def\opcit{{\it op.~cit.\/}}
\def\vs{vs.\hbox{}}
\def\areal{{\text{areal}}}
\def\bh{{\text{bh}}}

\def\ccenter{{\text{center}}}
\def\final{{\text{final}}}
\def\fit{{\text{fit}}}
\def\Gaussian{{\text{Gaussian}}}
\def\H{{\text{H}}}
\def\inner{{\text{inner}}}
\def\init{{\text{init}}}

\def\max{{\text{max}}}
\def\min{{\text{min}}}
\def\MS{{\text{MS}}}

\def\out{{\text{out}}}
\def\perturb{{\text{perturb}}}
\def\Schw{{\text{Schw}}}
\def\total{{\text{total}}}
\def\York{{\text{York}}}
\def\ab{_{ab}}
\def\ij{_{ij}}
\def\upij{^{ij}}
\def\dfrac#1#2{{\displaystyle \frac{#1}{#2}}}
\def\tfrac#1#2{{\textstyle \frac{#1}{#2}}}
\def\half{\frac{1}{2}}

\def\thalf{\tfrac{1}{2}}
\def\tthird{\tfrac{1}{3}}
\def\tquarter{\tfrac{1}{4}}
\def\csmash#1{\hbox to 0em{\hss{#1}\hss}}	
\def\cwso#1#2{{\setbox0=\hbox{#2}\hbox to\wd0{\hss{#1}\hss}}}
\def\lwso#1#2{{\setbox0=\hbox{#2}\hbox to\wd0{{#1}\hss}}}
\def\rwso#1#2{{\setbox0=\hbox{#2}\hbox to\wd0{\hss{#1}}}}
\def\three{{}^{(3)}}				
\def\four{{}^{(4)}}				
\def\gtsim{\gtrsim}
\def\ltsim{\lesssim}
\def\del{\nabla}
\def\diag{\mathop{\rm diag}\nolimits}		
\def\Lie{\hbox{\pounds}}
\def\sub{{\vphantom{X_i^i}}}
\def\norm#1{\left\| #1 \right\|}
\def\gbar{\bar{g}}
\def\rbar{\bar{r}}
\def\tbar{\bar{t}}
\def\xbar{\bar{x}}
\def\alphahat{\hat{\alpha}}
\def\betahat{\hat{\beta}}
\def\ghat{\hat{g}}
\def\xhat{\hat{x}}
\def\wr{w}				

\def\G{{\sf G}}
\def\GG{{\cal G}}

\def\JJ{\text{\bf J}}

\def\Y{{\sf Y}}
\def\YY{{\cal Y}}
\def\eqn#1{$(\text{#1})$}
\def\eqref#1{\eqn{\protect\ref{#1}}}
\def\citeprefix{}
\def\citenameprefix#1{}
\def\citenamesuffix{}
\def\citesprefix{}
\def\citesnamprefix#1{}

\def\citenumericvsname#1#2{#2}		
\def\supersqueezetable{\let\tabbodyfont\tiny}
\def\fakefootnotemark#1{$^{\text{#1}}$}
\begin{document}
\preprint{gr-qc/9801087}
\title{A $3+1$ Computational Scheme for Dynamic			\\
       Spherically Symmetric Black Hole Spacetimes		\\
       -- I: Initial Data}
\author{Jonathan Thornburg
	\thanks{Address for correspondence:
		$t \,{\leq}\, \text{1998 Aug 31} \Rightarrow$
			Box~8--7, Thetis Island,
			British Columbia, V0R~2Y0, Canada,
			E-mail \hbox{\tt thornbur@theory.physics.ubc.ca}
		; $t \,{\geq}\, \text{1998 Sept 1} \Rightarrow$
			Institut \fur{} Theoretische Physik,
			\Universitat{} Wien,
			Boltzmanngasse~5,
			A-1090 Wien,
			Austria
}}
\address{Department of Physics and Astronomy,
	    University of British Columbia			\\
	 Vancouver, British Columbia, V6T~1Z1, Canada}
\date{\today}
\maketitle
%
%
\begin{abstract}
This is the first in a series of papers describing a $3+1$ computational
scheme for the numerical simulation of dynamic spherically symmetric
black hole spacetimes.  In this paper we discuss the construction of
dynamic black hole initial data slices using York's conformal-decomposition
algorithm in its most general form, where no restrictions are placed
on $K$ (the trace of the extrinsic curvature) and hence the full
4-vector nonlinear York equations must be solved numerically.

To construct an initial data slice, we begin with a known black hole
slice (\eg{} a slice of Schwarzschild or Kerr spacetime), perturb
this via some \Ansatz{} (\eg{} the addition of a suitable Gaussian
to one of the coordinate components of the 3-metric, extrinsic
curvature, or matter field variables), apply the York decomposition
(using a further \Ansatz{} for the inner boundary conditions) to
project the perturbed field variables back into the constraint
hypersurface, and finally optionally apply a numerical 3-coordinate
transformation to restore any desired form for the spatial coordinates
(\eg{} an areal radial coordinate).

In comparison to other initial data algorithms, the key advantage
of this algorithm is its flexibility: $K$ is unrestricted, allowing
the use of whatever slicing is most suitable for (say) a time evolution.
This algorithm also offers great flexibility in controlling the
physical content of the initial data, while placing no restrictions
on the type of matter fields, or on spacetime's symmetries or lack
thereof.

We have implemented this algorithm for the spherically symmetric
scalar field system.  We present numerical results for a number of
asymptotically flat Eddington-Finkelstein--like initial data slices
containing black holes surrounded by scalar field shells, the latter
with masses ranging from as low as $0.17$ to as high as $17$~times
the black hole mass.  In all cases we find that the computed slices
are very accurate: Using 4th~order finite differencing on smoothly
nonuniform grids with resolutions of $\Delta r/r \approx 0.02$ ($0.01$)
near the perturbations, and Gaussian perturbations yielding scalar field
shells with relative width $\sigma/r \approx \tfrac{1}{6}$ and with about
$2/3$~the black hole mass, the numerically computed energy and momentum
constraints for the final slices are ${\ltsim}\, 10^{-8}$ ($10^{-9}$)
and ${\ltsim}\, 10^{-9}$ ($10^{-10}$) in magnitude respectively.  For
similarly perturbed numerically computed ``warped'' vacuum slices, the
Misner-Sharp mass function is independent of position, and the 4-Riemann
quadratic curvature invariant $R_{abcd} R^{abcd}$ is equal to its
(position-dependent) Schwarzschild-spacetime value, to within relative
errors of ${\ltsim}\, 10^{-5}$ ($5{\times}10^{-7}$) and
${\ltsim}\, 3{\times}10^{-4}$ ($5{\times}10^{-5}$) respectively.
All these errors show the expected $O \bigl( (\Delta r)^4 \bigr)$
scaling with grid resolution $\Delta r$.

Finally, we briefly discuss the errors incurred when interpolating
data from one grid to another, as in numerical coordinate
transformation or horizon finding.  We show that for the usual
moving-local-interpolation schemes, even for smooth functions the
interpolation error is {\em not smooth\/}.  If interpolated data
is to be differentiated, we argue that either the interpolation
order should be raised to compensate for the non-smoothness, or
explicitly smoothness-preserving interpolation schemes should be
used.
\end{abstract}
%
%
\draft				
%
%
%
%
\section{Introduction}
\label{sect-introduction}

Dynamic black hole spacetimes are interesting both astrophysically
(\eg{} in the study of supernova core collapse, active-galactic-nuclei
engines, and binary neutron-star or black-hole decay and coalescence)
and as laboratories in which to study the physics of strong-field
gravitation itself (\eg{} horizons, causality, cosmic censorship,
and self-similarity).  Such systems are time-dependent, strongly
relativistic, and often highly asymmetric, and so are hard to study
in detail except by numerical methods.

This is the first in a series of papers describing a $3+1$ computational
scheme for the numerical simulation of dynamic black hole spacetimes.
We discuss both our general computational scheme (which should be
applicable to a wide range of dynamic black hole spacetimes), and its
application to a specific test system, that of an asymptotically flat
spherically symmetric spacetime containing a black hole surrounded by
a (massless, minimally coupled) scalar field.
\footnote{
	 We have previously (\citeprefix\cite{Thornburg-PhD})
	 applied a similar computational scheme to a different
	 test system, that of an asymptotically flat axisymmetric
	 vacuum spacetime containing a black hole surrounded
	 by gravitational radiation.  Unfortunately, for that
	 system we have been unable to obtain time evolutions
	 free of finite differencing instabilities.
	 }
{}  Accordingly, we first work out our formalism and describe our
algorithms in generic terms (making no assumptions about spacetime
symmetries or what types of matter fields may be present), and only
then specialize first to the scalar field system, and finally to the
spherically symmetric scalar field system.

In this paper we focus on the \defn{initial data problem} of constructing
suitable initial data on a Cauchy surface.
\footnote{
	 The term \defn{initial value problem} is sometimes
	 used here.  We find this terminology undesirable
	 because it conflicts with standard usage elsewhere
	 (\eg{} in the study of differential equations, and
	 indeed in this journal's own Physics and Astronomy
	 Classification Scheme), where it refers to the
	 evolution of initial data, \ie{} the time integration
	 of a set of ODEs or PDEs given values of the state
	 variables at some initial time.
	 }
{}  That is, we consider the problem of computing the 3-metric and
extrinsic curvature for an initial data slice, such that the constraints
are satisfied and the slice has a desired physical content.

For black hole spacetimes, we have the additional problem of dealing
with singularities.  Traditionally this has been done using freezing
slicings (see, for example,
\citesprefix\cite{Smarr-1984-in-GR10,Shapiro-Teukolsky-1986-in-Centrella}),
but more recently attention has focused on the \defn{black hole exclusion}
or \defn{horizon boundary condition} technique
\footnote{
	 We prefer the former term as focusing
	 attention on the underlying process --
	 the exclusion of the black hole from the
	 computational domain -- rather than on
	 the particular boundary condition used
	 to implement it.
	 }
{} (see, for example,
\citesprefix\cite{Thornburg-MSc,Thornburg-1987-2BH-initial-data,
Thornburg-1991-BHE-talk,Seidel-Suen-1992-BHE,Thornburg-PhD,
Scheel-Shapiro-Teukolsky-1995a-BHE-Brans-Dicke,ADMSS-1995-BHE,
Scheel-Shapiro-Teukolsky-1995b-BHE-Brans-Dicke,Marsa-PhD,
Marsa-Choptuik-1996-sssf,SBCST-1997-3D-BHE-evolution}),
where the slices penetrate the event horizon and meet the
singularity or singularities, but in each slice the computational
domain (numerical grid) excludes or omits roughly the black hole
region.  (The precise region excluded varies between different
black-hole--exclusion computational schemes.)

The black hole exclusion technique requires special consideration
when constructing initial data, because the boundary of the excluded
region becomes a new \defn{inner} boundary for the computational
domain, typically on or just inside the (each) black hole's outermost
apparent horizon.
\footnote{
	 As discussed elsewhere
	 (\citesprefix\cite{Thornburg-PhD,Thornburg-1998-sssf-evolution}),
	 our particular form of the black hole
	 exclusion technique currently places the
	 inner boundary considerably inside the
	 apparent horizon (typically at ${\sim}\, 75\%$
	 of the horizon radius), but the details
	 of this are unimportant here.
	 }
{}  This new boundary requires corresponding boundary conditions
for the (elliptic) constraint equations.

Because the computational domain includes the (event) horizon, the
black hole exclusion technique requires that the slicing and the
$3+1$ field variables all be nonsingular and smooth there, and in
fact throughout some neighbourhood of the horizon.  This generally
rules out maximal slices such as the Schwarzschild (Boyer-Lindquist)
slices in Schwarzschild (Kerr) spacetimes; Eddington-Finkelstein (Kerr)
or similar slices are usually used instead.  For our purposes, the
key property of these latter slices is that they have $K$, the trace
of the extrinsic curvature, nonzero and spatially variable over most
or all of the slices.

Our slices having generic $K$ has an important impact on the initial
data construction itself:  York's conformal-decomposition technique
is widely used for solving the initial data problem, and we use it
here, but historically it's usually been used in one of several
special cases in which the York equations simplify greatly.  However,
as discussed in section~\ref{sect-special-cases-of-York-decomposition},
none of these special cases apply for slices with generic $K$, so we
must (numerically) solve the York equations in their full form.

In the remainder of this paper,
we first summarize our notation
	(section~\ref{sect-notation}),
then discuss the York conformal decomposition, its various special
cases, our general methods for solving its equations, and its outer
boundary conditions
	(sections~\ref{sect-York-decomposition}--\ref{sect-outer-BCs}).
We then describe our \Ansatze{} for choosing the inputs to and the
inner boundary conditions for the York decomposition
	(section~\ref{sect-initial-data-algorithm+ansatze}).
We next describe our diagnostics for assessing the accuracy of our
initial data and studying its physical content
	(section~\ref{sect-diagnostics}),
introduce our spherically-symmetric-scalar-field test system
	(section~\ref{sect-scalar-field+spherical-symmetry}),
describe our numerical methods
	(section~\ref{sect-numerical-methods}),
and present some sample results from a numerical code incorporating
these techniques
	(section~\ref{sect-sample-results}).
We end the main body of the paper with some conclusions and directions
for further research
	(section~\ref{sect-conclusions}).
In the appendices we tabulate some of the $3+1$ field variables for
an Eddington-Finkelstein slice of Schwarzschild spacetime
	(appendix~\ref{app-Schw-EF}),
tabulate the detailed equations for our
spherically-symmetric-scalar-field test system
	(appendix~\ref{app-sssf-eqns}),
discuss the derivation of our $3+1$ form of the Misner-Sharp mass
function
	(appendix~\ref{app-mass-function}),
describe our numerical-coordinate-transformation algorithm
	(appendix~\ref{app-numerical-coord-xforms}),
describe our methods for quantitatively testing finite differencing
convergence
	(appendix~\ref{app-convergence-tests}),
and discuss the non-smoothness of errors when interpolating data from
one grid to another
	(appendix~\ref{app-non-smoothness-of-interp-errors}).
%
%
\section{Notation}
\label{sect-notation}

We generally follow the sign and notation conventions of
\citenameprefix{Misner, Thorne, and Wheeler}
\citeprefix\cite{MTW}
\citenamesuffix{},
with a $(-,+,+,+)$ spacetime metric signature, $G = c = 1$ units,
and all masses and coordinate distances also taken as dimensionless.
We assume the usual Einstein summation convention for all repeated
indices, and we use the Penrose abstract-index notation, as described
by (\eg{}) \citeprefix\cite{Wald}.  However, for pedagogical convenience
we often blur the distinction between a tensor as an abstract geometrical
object and the vector or matrix of a tensor's coordinate components.

We use the standard $3+1$ formalism of \citeprefix\cite{ADM-1962}
(see \citesprefix\cite{York-1979-in-Yellow,York-1983-in-Red} for
recent reviews).  For our spherically-symmetric--scalar-field test
system, we use coordinates $(t,r,\theta,\phi)$, with $\theta$ and
$\phi$ the usual spherical-symmetry coordinates.  However, we leave
$t$ and $r$ arbitrary, \ie{} we make no assumptions about the choice
of the lapse function or (the radial component of) the shift vector.

The distinction between 3- and 4-tensors is usually clear from context,
but where ambiguity might arise we use prefixes $\three$ and $\four$
respectively, as in $\three\! R$ and $\four\! R$.  Any tensor without
a prefix is by default a 3-tensor.  $\Lie_v$ denotes the Lie derivative
operator along the 4-vector field $v^a$.  $\delta^i{}_j$ is the usual
Kronecker delta.

We use $abcd$ for spacetime (4-)\,indices, and $\partial_a$
denotes the spacetime coordinate partial derivative operator
$\partial / \partial x^a$.  $g\ab$ denotes the spacetime metric
and $\del_a$ the associated 4-covariant derivative operator.

We use $ijkl$ for spatial (3-)\,indices.  $\partial_i$ denotes
the spatial coordinate partial derivative operator
$\partial / \partial x^i$.  $g\ij$ denotes the 3-metric of a slice,
$\del_i$ the associated 3-covariant derivative operator, and $g$
the determinant of the matrix of $g\ij$'s coordinate components.
$\alpha$ and $\beta^i$ denote the $3+1$ lapse function and shift
vector respectively.  $n^a$ denotes the (timelike) future pointing
unit normal to the slices.
$K\ij \equiv \thalf \Lie_n g\ij \equiv - \del_i n_j$ denotes the
3-extrinsic curvature of a slice, and $K \equiv K_i{}^i$ its trace.
$\rho \equiv n^a n^b T\ab$ and $j^i = - n_a T^{ai}$ denote the locally
measured energy and 3-momentum densities respectively.  $T\ij$ denotes
the spatial stress-energy tensor, and $T = T_i{}^i$ its trace.

The $3+1$ energy and momentum constraints are thus
\begin{mathletters}
							\label{eqn-constraints}
\begin{eqnarray}
C \equiv
	\Bigl( R - K\ij K\upij + K^2 \Bigr)
	-
	\Bigl( 16 \pi \rho \Bigr)
		& = &	0
									\\
C^i \equiv
	\Bigl( \del_j K\upij - \del^i K \Bigr)
	-
	\Bigl( 8 \pi j^i \Bigr)
		& = &	0
\end{eqnarray}
\end{mathletters}
respectively, where $R$ denotes the 3-Ricci scalar computed from
$g\ij$ in the usual manner.

$\diag[ \cdots ]$ denotes the diagonal matrix with the specified
diagonal elements.  $\Gaussian(x{=}A, \sigma{=}B)$ denotes the Gaussian
function $\exp(-\thalf z^2)$, where $z \equiv (x - A) / B$.
%
%
\section{The York Decomposition}
\label{sect-York-decomposition}

\citenameprefix{York}
\citesprefix\cite{York-1971-York-decomposition,
York-1972-York-decomposition,York-1973-York-decomposition,
OMurchadha-York-1974-York-decomposition-I,
OMurchadha-York-1974-York-decomposition-II}
\citenamesuffix{}
\citenumericvsname{has}{have} given an elegant theoretical analysis
of the $3+1$ initial data problem, which both clarifies its physical
meaning and supplies a practical algorithm for its solution.
(\citeprefix\cite{Isenberg-PhD} also discusses several related
topics related to the canonical formulation of gravity.)  This
\defn{conformal decomposition} approach to the initial data problem
is reviewed by \citesprefix\cite{York-1979-in-Yellow,
York-1983-in-Red,York-1985-in-Centrella-LeBlank-Bowers,
York-Piran-1982-in-Schild-lectures}.  \citeprefix\cite{Choptuik-MSc}
also gives a very clear presentation of the York decomposition.

For present purposes, we view the York decomposition as defining a
projection operator $\YY$ which projects the field variables into
the constraint hypersurface.  In detail, $\YY$ is defined as follows:

Given an initial (\defn{base}) 3-metric $g\ij$, extrinsic curvature
$K\ij$, and energy and 3-momentum densities $\rho$ and $j^i$, we
first split $g\ij$ into its determinant $g$ and the conformal
metric $h\ij \equiv g^{-1/3} g\ij$, and $K\ij$ into its trace
$K$ and the tracefree extrinsic curvature
$E\ij \equiv K\ij - \tthird g\ij K$.  We then define the
covariant derivative $\del_i$ and compute the 3-Ricci tensor
and scalar $R\ij$ and $R$, all using the base metric $g\ij$
in the usual manner.

Next, we solve the nonlinear elliptic \defn{York equation}
\begin{mathletters}
							\label{eqn-York}
\begin{eqnarray}
\del_i \del^i \Psi
	& = &	\tfrac{1}{8} R \Psi
		+ \tfrac{1}{12} K^2 \Psi^5
		- \tfrac{1}{8} F\ij F\upij \Psi^{-7}
		- 2 \pi \rho \Psi^{-3}
					\label{eqn-York-energy-constraint}
									\\
{(\Delta_\ell \Omega)}^i
	& = &	\tfrac{2}{3} (\del^i K) \Psi^6
		- \del_j E\upij
		+ 8 \pi j^i
					\label{eqn-York-momentum-constraint}
\end{eqnarray}
\end{mathletters}
for the 4-vector
\footnote{
	 Here and throughout our discussion
	 of the York decomposition and equations,
	 we use the term ``4-vector'' only in
	 the sense of ``4-tuple of real-valued
	 functions'', not in the sense of
	 ``rank~1 4-tensor''.
	 }
{} $\Y \equiv (\Psi,\Omega^i)$, where
$F\upij \equiv E\upij + {(\ell \Omega)}\upij$, and where the York
\defn{vector divergence} and \defn{vector Laplacian} operators $\ell$
and $\Delta_\ell$ are defined by
\begin{mathletters}
\begin{eqnarray}
{(\ell X)}\upij
	& = &	\del^i X^j + \del^j X^i - \tfrac{2}{3} g\upij \del_k X^k
									\\
{(\Delta_\ell X)}^i
	& = &	\del_j {(\ell X)}\upij
									\\
	& = &	\del_j \del^j X^i
		+ \tthird \del^i \del_j X^j
		+ R^i{}_j X^j
\end{eqnarray}
\end{mathletters}
for any (arbitrary) contravariant 3-vector $X^i$.  We discuss
boundary conditions for the York equation in sections~\ref{sect-outer-BCs}
and~\ref{sect-initial-data-algorithm+ansatze}.

Finally, we define the output field variables
$(g\ij, K\ij, \rho, j^i)\sub_\out \equiv \YY (g\ij, K\ij, \rho, j^i)$
by
\begin{mathletters}
							\label{eqn-York-xform}
\begin{eqnarray}
(g\ij)\sub_\out
	& = &	\Psi^4 g\ij
									\\
(K\ij)\sub_\out
	& = &	\Psi^{-2} F\ij + \tfrac{1}{3} K \Psi^4 g\ij
									\\
(\rho)\sub_\out
	& = &	\Psi^{-8} \rho
									\\
(j^i)\sub_\out
	& = &	\Psi^{-10} j^i
\end{eqnarray}
\end{mathletters}

Although it's not obvious from the form of the equations, the
output field variables determined in this way do indeed satisfy
the constraints.
%
%
\section{Special Cases of the York Decomposition}
\label{sect-special-cases-of-York-decomposition}

As discussed by \citesprefix\cite{York-1979-in-Yellow,
Bowen-York-1980-2BH-initial-data,York-Piran-1982-in-Schild-lectures,
York-1983-in-Red,York-1989-in-Frontiers} and reviewed by
\citeprefix\cite{Thornburg-PhD}, section~A5.1, the York decomposition
has several important special cases in which the general York
equations~\eqref{eqn-York} simplify considerably:
\begin{itemize}
\item	If the base field variables define a \xdefn{maximal slice}
	($K = 0$), or more generally if $K$ is spatially constant,
	then the 4-vector nonlinear elliptic system~\eqref{eqn-York}
	decouples into separate {\em linear\/} 3-vector (momentum) and
	nonlinear scalar (energy) constraint equations, which may be
	solved sequentially.
\item	If the base field variables define a slice which is both
	maximal and \xdefn{3-conformally flat} ($g\ij = \Phi^4 f\ij$
	for some conformal factor $\Phi$ and flat 3-metric $f\ij$),
	the York equations simplify still further.  In particular,
	for this case
\citenameprefix{Bowen}
\citesprefix\cite{Bowen-PhD,
Bowen-1979-momentum-constraint-analytical-soln,
Bowen-York-1980-2BH-initial-data,
Bowen-1982-momentum-constraint-analytical-soln}
\citenamesuffix{}
	\citenumericvsname{has}{have} found the (analytical)
	general solution for the (linear) momentum constraint
	equation~\eqref{eqn-York-momentum-constraint} subject
	to outer boundary conditions appropriate for an
	asymptotically flat spacetime.  This means that
	only the (nonlinear) energy constraint
	equation~\eqref{eqn-York-energy-constraint} need be solved
	numerically.  Bowen (\opcit{}) has also found several
	\Ansatze{} for choosing the base extrinsic curvature $K\ij$
	in this case, such that the final field variables represent
	an initial slice containing a single black hole with freely
	specifiable mass, momentum, and spin.  Since the momentum
	constraint equation~\eqref{eqn-York-momentum-constraint}
	is linear here, these base $K\ij$ may be superimposed to
	obtain base $K\ij$ values for multiple--black-hole slices.
	We discuss this case further below.
\item	If the base field variables define a slice which is
	\xdefn{time-symmetric} ($K\ij = 0$) and 3-conformally
	flat, then the full 4-vector York equation~\eqref{eqn-York}
	has a fully-analytical asymptotically flat solution,
	first explicitly found by
\citenameprefix{Misner}
\citeprefix\cite{Misner-1960-initial-data}
\citenamesuffix{},
	following an earlier suggestion of
\citenameprefix{Einstein and Rosen}
\citeprefix\cite{Einstein-Rosen-1935-bridge}
\citenamesuffix{}.
	The Misner solution, first studied numerically by
	\citeprefix\cite{Brill-Lindquist-1963-Misner-initial-data-energy},
	represents an arbitrary number of momentarily stationary
	``generalized Schwarzschild'' black holes, each with
	freely specifiable ``mass''.
\fakefootnotemark{\ref{footnote-hard-to-define-mass-etal-in-N-BH-slice}}
\end{itemize}

The special case where the initial slice is maximal and
3-conformally flat has been used in most past numerical work on
the initial data problem.  In particular, several authors, notably
\citesprefix\cite{York-Piran-1982-in-Schild-lectures,Choptuik-MSc,
Dubal-1992-3D-BH-initial-data}, have used this technique to construct
initial data slices containing single black holes with varying spins,
surrounded by various amounts of gravitational radiation.

There have also been several extensions of this work to the
multiple black hole case.
\citesprefix\cite{Kulkarni-Shepley-York-1983-N-BH-initial-data,
Kulkarni-1984-N-BH-initial-data,York-1984-N-BH-initial-data,
Bowen-1985-N-BH-initial-data}
have used infinite series of ``image charges'' to construct a
semianalytical solution for maximal 3-conformally flat base field
variables satisfying \defn{reflection symmetric} or \defn{inversion
symmetric} inner boundary conditions on the surfaces of each of
$N$~black holes, and
\citesprefix\cite{Bowen-York-1980-2BH-initial-data,
Bowen-Rauber-York-1984-2BH-initial-data,Kulkarni-1984-2BH-initial-data}
have specialized this approach to the $2$-black-hole case.
Using these latter solutions, various authors, for example
\citesprefix\cite{Rauber-PhD,Rauber-1986-in-Centrella,Cook-PhD,
Cook-1991-2BH-initial-data,Dubal-Oliveira-Matzner-1992-in-dInverno,
CCDKMO-1993-3D-2BH-initial-data,
Matzner-Huq-Shoemaker-1998-N-BH-initial-data},
have constructed a variety of numerical solutions for $\Psi$
corresponding to initial data slices containing two black holes, each
with freely specifiable
\footnote{
	 The question of just what can and
	 can't be freely specified for these
	 initial slices is somewhat subtle.
	 \citeprefix\cite{York-1989-in-Frontiers}
	 gives a brief discussion of this;
	 see also
\citeprefix\cite{Matzner-Huq-Shoemaker-1998-N-BH-initial-data}.
	 }
{} ``mass'', ``momentum'', and ``spin''.
\footnote{
\label{footnote-hard-to-define-mass-etal-in-N-BH-slice}
	 ``Mass'', ``momentum'', and ``spin'' are
	 in quotes here because (outside of spherical
	 symmetry) these quantities aren't uniquely
	 defined for the individual black holes,
	 unless they're far enough apart to have
	 separate approximate asymptotically flat
	 regions.
	 }
{}  \citesprefix\cite{Thornburg-MSc,Thornburg-1987-2BH-initial-data}
\citenumericvsname{have}{has} also constructed maximal 3-conformally-flat
2-black-hole numerical solutions for $\Psi$, but using an apparent
horizon boundary condition (a version of the black hole exclusion
technique) instead of reflection symmetry.

\citeprefix\cite{Oohara-Nakamura-1989-in-Frontiers}
\citenumericvsname{has}{have} used a modified version of Bowen's
analytical solution for $\Omega^i$ to construct initial data slices
modeling a binary neutron star system.  Because no black holes are
present in this case, they avoided the problem of inner boundary
conditions altogether.

As noted in the introduction, however, in this paper our concern is
with initial slices where $K$ is nonzero and spatially variable, so
none of the special cases discussed in this section apply, \ie{} we
must solve the full 4-vector York equation~\eqref{eqn-York} numerically.
%
%
\section{Solving the York Equations}
\label{sect-solving-York-eqns}

The York equation~\eqref{eqn-York} is a nonlinear elliptic PDE.
To solve it, we first use a global Newton-Kantorovich ``outer''
iteration (\citeprefix\cite{Boyd}, appendix~C), then use standard
finite differencing methods to numerically solve each of the resulting
sequence of ``inner'' linear elliptic PDEs.

In detail, we first rewrite the York equation~\eqref{eqn-York} in
the form
\begin{mathletters}
							\label{eqn-York-G}
\begin{eqnarray}
\G^0 \equiv
	\del_i \del^i \Psi
	- \tfrac{1}{8} R \Psi
	- \tfrac{1}{12} K^2 \Psi^5
	+ \tfrac{1}{8} F\ij F\upij \Psi^{-7}
	+ 2 \pi \rho \Psi^{-3}
		& = &	0
									\\
\G^i \equiv
	{(\Delta_\ell \Omega)}^i
	- \tfrac{2}{3} (\del^i K) \Psi^6
	+ \del_j E\upij
	- 8 \pi j^i
		& = &	0
\end{eqnarray}
\end{mathletters}
where we define $\G \equiv \GG(\Y)$ to be the left hand side 4-vector.
We then linearize the continuum differential operator $\GG$ about
the current continuum solution estimate $\Y$,
\begin{equation}
\GG(\Y + \delta\Y)
	= \GG(\Y)
	  + \JJ \Bigl[ \GG(\Y) \Bigr] (\delta\Y)
	  + O \bigl( \norm{\delta\Y}^2 \bigr)
						\label{eqn-York-linearized}
\end{equation}
where $\delta\Y$ is a finite perturbation in $\Y$ and the linear
differential operator $\JJ[\GG(\Y)]$ is the linearization of the
differential operator $\GG$ about the point $\Y$.  We then neglect
the higher-order (nonlinear) terms in~\eqref{eqn-York-linearized}
and solve for the perturbation $\delta\Y$ such that
$\GG(\Y + \delta\Y) = 0$.  This gives the linear elliptic PDE
\begin{equation}
\JJ \Bigl[ \GG(\Y) \Bigr] (\delta\Y) = - \G
							\label{eqn-York-update}
\end{equation}
to be solved for $\delta\Y$.  Finally, we update the approximate
solution via $\Y \to \Y + \delta\Y$, and repeat the iteration until
$\norm{\G}$ is small.

It's convenient to rewrite the definition~\eqref{eqn-York-linearized}
of the Jacobian operator $\JJ$ in the differential form
$d\G = \JJ(d\Y)$ to more clearly express $\JJ$'s physical meaning of
mapping infinitesimal changes in $\Y$ to infinitesimal changes in
$\G$.  With this interpretation in mind, it's easy to see
from~\eqref{eqn-York-G} that the Jacobian is given by
\begin{mathletters}
						\label{eqn-York-G-Jacobian}
\begin{eqnarray}
d\G^0	& = &	\del_i \del^i \, d\Psi
		- \tfrac{1}{8} R \, d\Psi
		- \tfrac{5}{12} K^2 \Psi^4 \, d\Psi
								\nonumber
									\\
	&   &	{}
		- \tfrac{7}{8} F\ij F\upij \Psi^{-8} \, d\Psi
		+ \tfrac{1}{4} F\ij \Psi^{-7} {(\ell \, d\Omega)}\upij
		- 6 \pi \rho \Psi^{-4} \, d\Psi
									\\
d\G^i	& = &	- 4 (\del^i K) \Psi^5 \, d\Psi
		+ {(\Delta_\ell \, d\Omega)}^i
\end{eqnarray}
\end{mathletters}

Given a reasonable numerical solution of the updating
equation~\eqref{eqn-York-update} (discussed in
section~\ref{sect-numerical-methods}), we find in practice that
the Newton-Kantorovich iteration converges rapidly and robustly,
typically reducing $\norm{\G}_\infty$ to negligible levels
(\eg{} ${<}\, 10^{-10}$) in $3$--$5$ iterations.
%
%
\section{Outer Boundary Conditions}
\label{sect-outer-BCs}

The York equation~\eqref{eqn-York} is an elliptic PDE, and as such
needs boundary conditions.  Assuming the black hole exclusion technique,
there are $N+1$ boundaries for a slice with $N$ black holes: an
inner boundary near each black hole's horizon, and an outer boundary
near spatial infinity.  We discuss the inner boundary conditions in
section~\ref{sect-initial-data-algorithm+ansatze}.

The outer boundary conditions are determined by the desired
asymptotic forms of $\Psi$ and $\Omega^i$ at spatial infinity.  For
an asymptotically flat spacetime, we have
\begin{mathletters}
					\label{eqn-outer-BC-at-infinity}
\begin{eqnarray}
\Psi	& = &	1
				\qquad
				\text{(at spatial infinity)}
									\\
\Omega^i
	& = &	0
				\qquad
				\text{(at spatial infinity)}
\end{eqnarray}
\end{mathletters}

Purely for the initial data computation (with no time evolution
involved), one can treat the unbounded problem domain by compactifying,
\eg{} by introducing a new radial coordinate $s \equiv 1 - r_\min/r$.
This has been done by, for example, \citesprefix\cite{Choptuik-MSc,
Rauber-PhD,Rauber-1986-in-Centrella}.

However, if the initial data computation is being done in conjunction
with a $3+1$ time evolution code (as is our main interest here), it's
typically more convenient to not compactify the domain, but instead
apply an approximate outer boundary condition at some finite outer
boundary radius $r_\max$.  When doing this, we can exploit the known
asymptotic falloff of $\Psi$ and $\Omega^i$ to obtain more accurate
results than would be obtained by simply applying the spatial-infinity
condition~\eqref{eqn-outer-BC-at-infinity} at $r = r_\max$.  As
discussed by \citesprefix\cite{York-1979-in-Yellow,
York-1980-in-Taub-festschrift,York-Piran-1982-in-Schild-lectures,
York-1983-in-Red,Oohara-Nakamura-1989-in-Frontiers,
York-1989-in-Frontiers,Cook-PhD,Cook-1991-2BH-initial-data,
OMurchadha-1992-in-dInverno}, this gives the Robin outer boundary
conditions
\begin{mathletters}
\begin{eqnarray}
(\del_k \Psi) n^k + \frac{\Psi - 1}{r}
	& = &	O \left( \frac{1}{r^3} \right)
		\approx	0
				\qquad
				\text{(at $r = r_\max$)}
					\label{eqn-Psi-outer-BC-scalar-Robin}
									\\
n_j
\left( \delta^i{}_k - \tfrac{1}{2} n^i n_k \right)
{(\ell \Omega)}^{kj}
+ \frac{6}{7r}
  \left( \delta^i{}_k - \tfrac{1}{8} n^i n_k \right)
  \Omega^k
        & = &	O \left( \frac{1}{r^3} \right)
		\approx	0
				\qquad
				\text{(at $r = r_\max$)}
					\label{eqn-Omega-outer-BC-vector-Robin}
\end{eqnarray}
\end{mathletters}

[In practice, one can often still obtain adequate accuracy by replacing
the vector Robin condition~\eqref{eqn-Omega-outer-BC-vector-Robin} by
the simple (and tensorially incorrect) application of a scalar Robin
condition (analogous to~\eqref{eqn-Psi-outer-BC-scalar-Robin}) separately
to each coordinate component of $\Omega^i$,
\begin{equation}
(\del_k \Omega^i) n^k + \frac{\Omega^i}{r}
	=	O \left( \frac{1}{r^3} \right)
		\approx	0
				\qquad
				\text{(at $r = r_\max$)}
					\label{eqn-Omega-outer-BC-scalar-Robin}
\end{equation}
We use this approximation in our numerical code, and find that it
gives good results.]

[It's also useful to note that for many coordinate conditions, the
asymptotic forms of $\Psi$ and $\Omega^i$ are identical in their
leading terms to those for the $3+1$ lapse function $\alpha$ and
shift vector $\beta^i$ (respectively).  Hence for such coordinate
conditions the boundary conditions for $\Psi$ and $\Omega^i$ are
also identical to those for $\alpha$ and $\beta^i$.  In the context
of a $3+1$ code incorporating both initial data computation and
time evolution, this may allow the same outer boundary condition
code to be used for the York equation as for the coordinate conditions.]
%
%
%
\section{Our Initial Data Algorithm and \ANSATZE{}}
\label{sect-initial-data-algorithm+ansatze}

Given the York projection operator, our overall initial data algorithm
is as follows:
\begin{enumerate}
\item	\label{algorithm-step-analytical-BH-slice}
	We begin with an exact (analytically known) black hole
	initial data slice, chosen according to some \Ansatz{}.
	As discussed in sections~\ref{sect-introduction}
	and~\ref{sect-special-cases-of-York-decomposition},
	this will typically have $K$ nonzero and spatially variable
	throughout the slice.
\item	\label{algorithm-step-perturbation}
	We apply an arbitrary (in general constraint-violating)
	perturbation, again chosen according to some \Ansatz{}.
\item	\label{algorithm-step-York-projection}
	We apply the York decomposition to project the perturbed
	field variables back into the constraint hypersurface,
	solving the full 4-vector York equations numerically.
	Here the outer boundary conditions are dictated by the
	physics (\cf{}~section~\ref{sect-outer-BCs}), but an
	(another) \Ansatz{} is needed to supply the inner
	boundary conditions.
\item	\label{algorithm-step-coord-xform}
	Optionally, we apply a numerical 3-coordinate transformation
	to restore the spatial coordinates within the slice to some
	desired form, \eg{} to make the radial coordinate areal
	($g_{\theta\theta} = r^2$).  We discuss numerical coordinate
	transformations in appendix~\ref{app-numerical-coord-xforms}.
\end{enumerate}

In comparison to other initial data algorithms, the key advantage
of using the full York decomposition is its flexibility:  Since no
restriction is placed on $K$, almost any slicing may be used.  That
is, with this algorithm the slicing may be chosen based on its
suitability for (say) a black-hole--excluded time evolution, rather
than being limited by the initial data solver.  As well, the physical
content of the initial slice may be controlled over a wide range by
varying the choices for the different \Ansatze{}.  This algorithm
also places no restrictions on what type(s) of matter field(s) may
be present, and (given suitable programming for the numerical solution
of the York equations) it works equally well in spherical symmetry,
in axisymmetry, or in fully generic spacetimes with no continuous
symmetries.

To actually use this algorithm to construct initial data slices,
we must make suitable choices for the various \Ansatze{}.  We now
consider these choices:

As noted in the introduction, the black hole exclusion technique
requires a slicing which penetrates the horizon, with the slicing
and the $3+1$ field variables all nonsingular and smooth throughout
some neighbourhood of the horizon.  We thus use an Eddington-Finkelstein
slice of Schwarzschild spacetime, or a Kerr slice of Kerr spacetime,
\footnote{
	 Recall that these slices are both defined
	 (\citeprefix\cite{MTW}, boxes~31.2 and~33.2)
	 by taking the usual areal radial coordinate
	 $r$, and defining $t$ such that $t+r$ is
	 an ingoing null coordinate.  Both slices
	 are nonsingular near to and on the horizon.
	 For reference, in appendix~\ref{app-Schw-EF}
	 we tabulate some of the $3+1$ field variables
	 for an Eddington-Finkelstein slice of
	 Schwarzschild spacetime.
	 }
{} for step~\ref{algorithm-step-analytical-BH-slice} of our algorithm.

The choice of the perturbation \Ansatz{} for
step~\ref{algorithm-step-perturbation} of our algorithm is much
less clear-cut, and we have experimented with a number of different
choices for what field variable to perturb and what perturbation
to apply.  Provided the perturbation isn't too large, we have found
that in practice essentially any smooth localized perturbation
\footnote{
	 We haven't yet investigated
	 non-localized perturbations.
	 }
{} in the field variables ``works'', in the sense that the York
equations remain numerically well-behaved and yield a properly-constrained
initial data set (still) containing a black hole.  However, the
precise relationship between the form of the perturbation and the
nature of the resulting initial data slice is complicated and not
(yet) easily predicted a~priori.  In section~\ref{sect-sample-results}
we present numerical results for several examples of this relationship
(different choices of the perturbation), but much more research is
needed to explore and map this relationship's phenomenology.

To motivate the choice of an inner boundary condition
\Ansatz{} for the York equation~\eqref{eqn-York} in
step~\ref{algorithm-step-York-projection} of our algorithm, we
first observe that if the base field variables already satisfy
the constraints, then the York projection operator is the identity
operator, \ie{} the solution of the York equation~\eqref{eqn-York}
is $\Psi \equiv 1$ and $\Omega^i \equiv 0$.  We then consider the
slightly more general case where the base field variables are a
small perturbation off the constraint hypersurface.  To the extent
that the perturbation is indeed small, or at least that its effects
are small near the inner boundary, then $\Psi \equiv 1$ and
$\Omega^i \equiv 0$ should still be an approximate solution of the
York equation at the inner boundary.  This suggests taking
\begin{mathletters}
						\label{eqn-York-inner-BC}
\begin{eqnarray}
\Psi
	& = &	1
				\qquad
				\text{(at the inner boundary)}
									\\
\Omega^i
	& = &	0
				\qquad
				\text{(at the inner boundary)}
\end{eqnarray}
\end{mathletters}
as our inner boundary condition \Ansatz{} for the York
equation~\eqref{eqn-York}.

It's not a~priori obvious that (i)~this boundary condition is consistent
with any nontrivial solutions of the York equation~\eqref{eqn-York},
nor that (ii)~the resulting slices will (still) contain black holes.
However, from our numerical results in section~\ref{sect-sample-results},
both~(i) and~(ii) do in fact hold in practice.

Moreover, although our motivation for this boundary condition took
the perturbation to either be small, or at least to be small in its
effects near the inner boundary, our numerical
results in section~\ref{sect-sample-results/inner-BCs} also show
that this restriction isn't needed, \ie{} that accurate results
(initial data slices which satisfy the constraints) are still obtained
even for perturbations which are large and/or have large effects near
the inner boundary.  In fact, it appears that essentially {\em any\/}
reasonable inner boundary condition still yields accurate results.

We have previously (\citeprefix\cite{Thornburg-PhD}, sections~5.3
and~9.8) suggested that the fact that the physics of the initial data
problem doesn't specify any particular inner boundary condition
for the York equation, constitutes a serious drawback of the
overall black-hole--exclusion technique, and that the
\Ansatz{}~\eqref{eqn-York-inner-BC} therefore ``lacks a clear
physical motivation and is at best an ad-hoc solution''.  However,
we now regard this criticism as unwarranted:  Physically, the inner
boundary condition for the York equation specifies what type of
black hole (or more generally, spacetime region, \cf{}~our discussion
of the pqw5c slices in
section~\ref{sect-sample-results/other-perturbations}) is contained
within the inner boundary.  This {\em must\/} be specified as an
input to the initial data algorithm, and the
\Ansatz{}~\eqref{eqn-York-inner-BC} seems quite
reasonable for this purpose.

We have made a few limited experiments with inner boundary conditions
other than the simple Dirichlet condition~\eqref{eqn-York-inner-BC},
but much further research is needed to understand the relative merits
of different inner boundary conditions, and their effects on the
resulting initial data slices.
%
%
\section{Diagnostics}
\label{sect-diagnostics}

The purpose of any initial data algorithm is to construct slices
which satisfy the constraints, so we use the numerically computed
values of the constraints~\eqref{eqn-constraints}, (\ie{} their
deviations~$C$ and~$C^i$ from satisfaction), as diagnostics of our
initial slices' accuracy.  Because of its dependence through $R$ on
the 2nd~derivatives of the 3-metric components, the energy constraint
$C$ is particularly sensitive (useful) for this purpose; in practice,
nonzero computed values of $C$ primarily measure errors in the
computation of $R$.

We use a number of diagnostics to study the physical content of
our slices and the spacetimes in which they're embedded, including
the coordinate position(s) of any apparent horizon(s) in the slice,
the mass distribution in each slice (measured in spherical symmetry
by the Misner-Sharp mass function $m_\MS$ or the integrated scalar
field mass function $m_\mu$ defined below), the scalar field's
3-energy density $\rho$ (or the radial scalar field density
$4 \pi g_{\theta\theta} \rho$ in spherical symmetry), $K$, $R$, the
3-Ricci quadratic curvature invariant $R_{ij} R^{ij}$, the 4-Ricci
scalar $\four\! R$, and the 4-Riemann quadratic curvature invariant
$R_{abcd} R^{abcd}$.

As discussed by
\citesprefix\cite{Guven-OMurchadha-1995-constraints-in-spherical-symmetry-I},
in spherical symmetry the Misner-Sharp mass function $m_\MS$ can be
written as the volume integral of a local function
$\mu = \mu(\rho,j^i;g\ij,K\ij)$ of the scalar field variables.
We can thus define an alternative mass function (which should be
equal to the Misner-Sharp one) by volume-integrating $\mu$,
\begin{equation}
m_\mu = m_\MS(r_\min)
	+ \int _{r_\min} ^r
	       4 \pi g_{\theta\theta} \,
	       \mu \,
	       \sqrt{g_{rr}} \, dr
						\label{eqn-mass-function-mu}
\end{equation}
where $m_\MS(r_\min)$ is the Misner-Sharp mass function at the inner
grid boundary $r = r_\min$.  $m_\MS$ and $m_\mu$ are computed
independently and in very different ways (\eqref{eqn-mass-function-MS}
\vs{} \eqref{eqn-mu} and~\eqref{eqn-mass-function-mu(wr)}), so their
numerical equality at all radii is a useful consistency check and
accuracy diagnostic.  As discussed in
appendix~\ref{app-sssf-eqns/constraints+mass-fns+4-invariants}, for
our computational scheme $m_\mu$ is numerically better-behaved than
$m_\MS$, so we generally use $m_\mu$ as a general-purpose mass function,
and hereinafter we take $m$ (with no subscripts) to denote $m_\mu$.

By contracting the spacetime Einstein equations, it's easy to see
that $\four\! R = 8 \pi (\rho - T)$, so $\four\! R$ plays the role
of a 4-invariant diagnostic for the scalar field density.

$R_{abcd} R^{abcd}$ is a 4-invariant diagnostic for spacetime
curvature.  As discussed by \citeprefix\cite{York-1989-in-Frontiers},
$R_{abcd} R^{abcd}$ is given in terms of the $3+1$ variables by
\begin{eqnarray}
R_{abcd} R^{abcd}
	& = &
		8 R_{ij} R^{ij}					
		- 16 R_{ij} (K^i{}_k K^{kj} - K K^{ij})		
		+ 8 K_{ij} K^{jk} K_{kl} K^{li}			
								\nonumber
									\\
	&   &	{}
		- 16 K K_{ij} K^i{}_k K^{kj}			
		+ 8 K^2 K_{ij} K^{ij}				
		- 8 \del_{[i} K_{j]k} \, \del^i K^{jk}		
								\nonumber
									\\
	&   &	{}
		+ 16 \pi \Bigl[
			 4 T_{ij} (K^i{}_k K^{kj} - K K^{ij})	
			 - 4 R_{ij} T^{ij}			
			 \Bigr]
								\nonumber
									\\
	&   &	{}
		+ (16 \pi)^2 \Bigl[
			     T_{ij} T^{ij}			
			     - \tfrac{1}{4} T^2			
			     - \tfrac{9}{4} \rho^2		
			     + \tfrac{3}{2} T \rho		
			     \Bigr]
						\label{eqn-four-Riem-Riem}
\end{eqnarray}

Most of our diagnostics vary rapidly with spatial position
in any black hole spacetime, even Schwarzschild spacetime
(\cf~appendix~\ref{app-Schw-EF}).  For example, in an
Eddington-Finkelstein slice of a Schwarzschild spacetime, $R$
typically varies as ${\sim}\, 1/r^4$, where $r$ is the areal radius.
To focus on the deviations away from a
Schwarzschild\,/\,Eddington-Finkelstein slice, we usually normalize
out the ``background'' Schwarzschild variation by using \defn{relative}
diagnostics, \eg{} $R / R_\Schw$.  Here for any diagnostic $Z$, at
each areal radius, $Z_\Schw$ denotes the value of $Z$ at that same
areal radius in an Eddington-Finkelstein slice of a unit-mass
Schwarzschild spacetime, and more generally, for any fixed $m_\ast$,
$Z_{\Schw(m_\ast)}$ denotes the value of $Z$ at that same areal radius
in an Eddington-Finkelstein slice of a mass-$m_\ast$ Schwarzschild
spacetime.

When considering a (spherically symmetric) slice or region of a slice
whose mass function varies significantly with position, we also use
\defn{mass-relative} diagnostics, \eg{} $R / R_{\Schw(m(r))}$.  Here
for any slice $\cal S$ and diagnostic $Z$, at each areal radius~$r$,
$Z_{\Schw(m(r))}$ denotes the value of $Z$ at that same areal radius
$r$ in an Eddington-Finkelstein slice of a Schwarzschild spacetime of
mass $m(r)$, where $m(r)$ is the mass function of the slice $\cal S$
at the areal radius $r$.
\citesprefix\cite{Marsa-PhD,Marsa-Choptuik-1996-sssf} have used
the same notion of mass-relative variables in their initial data
construction, though not with diagnostics.
%
%
\section{Scalar Field and Spherical Symmetry}
\label{sect-scalar-field+spherical-symmetry}

We have previously (\citeprefix\cite{Thornburg-PhD}, appendix~5)
described the use of the same York-decomposition initial data problem
algorithm and \Ansatze{} described here, in constructing dynamic vacuum
axisymmetric initial data slices containing a black hole surrounded
by gravitational radiation.  Here we discuss the algorithm and
\Ansatze{}'s use in a different system, the construction of dynamic
spherically symmetric initial data slices (each) containing a black
hole surrounded by one or more scalar field shells.

The spherically symmetric scalar field and similar systems have been
studied by a number of past researchers, including (among others)
analytical studies by
\citesprefix\cite{Christoudolou-1986a-sssf,Christoudolou-1986b-sssf,
Christoudolou-1987a-sssf,Christoudolou-1987b-sssf,
Christoudolou-1991-sssf,Christoudolou-1993-sssf},
$3+1$ studies by
\citesprefix\cite{Choptuik-PhD,Choptuik-1991-consistency,
Seidel-Suen-1992-BHE,Bernstein-Bartnik-1995-sssf,
Scheel-Shapiro-Teukolsky-1995a-BHE-Brans-Dicke,
Scheel-Shapiro-Teukolsky-1995b-BHE-Brans-Dicke,ADMSS-1995-BHE,
Marsa-PhD,Marsa-Choptuik-1996-sssf},
$2+2$ studies by
\citesprefix\cite{Goldwirth-Piran-1987-sssf-2+2,
Goldwirth-Ori-Piran-1989-in-Frontiers,Hamade-Stewart-1996-sssf-2+2},
a hybrid $3+1$ and $2+2$ study by
\citeprefix\cite{GLPW-1996-sssf-3+1-and-2+2},
and an interesting comparison of $3+1$ and $2+2$ methods by
\citeprefix\cite{Choptuik-Goldwirth-Piran-1992-sssf-cmp-3+1-vs-2+2}.

[However, these authors haven't used the York decomposition for
constructing their initial data:  Focusing on the $3+1$ studies,
\citesprefix\cite{Choptuik-PhD,Choptuik-1991-consistency,
Bernstein-Bartnik-1995-sssf,
Scheel-Shapiro-Teukolsky-1995a-BHE-Brans-Dicke,
Scheel-Shapiro-Teukolsky-1995b-BHE-Brans-Dicke} all constructed
initial data using slicing conditions which simplified the constraints
sufficiently to allow a ``direct'' (algebraic) solution of the
constraints by suitably ordering the $3+1$ equations.
Using similar Eddington-Finkelstein--like slices to ours,
\citesprefix\cite{Marsa-PhD,Marsa-Choptuik-1996-sssf} used an
\Ansatz{} of setting $K^r{}_r = (K^r{}_r)\sub_{\Schw(m(r))}$,
then used a functional iteration scheme to satisfy the remaining
constraint equations.
\citesprefix\cite{Seidel-Suen-1992-BHE,ADMSS-1995-BHE} used
Schwarzschild spacetime as their only general relativistic
system, and hence needed only to use an appropriate slice of
Schwarzschild spacetime for their initial data.]

Turning now to our scalar field formalism, we begin with the generic
case, making no assumptions about about any spacetime symmetries.
Following \citesprefix\cite{Choptuik-PhD,Choptuik-1991-consistency,
Marsa-PhD,Marsa-Choptuik-1996-sssf}, we take the scalar field $\phi$
to satisfy the 4-scalar wave equation $\del_a \del^a \phi = 0$,
and to have the stress-energy tensor
\begin{equation}
4 \pi T\ab = (\partial_a \phi) (\partial_b \phi)
	     - \thalf g_{ab} (\partial_c \phi) (\partial^c \phi)
\end{equation}
We then define the $3+1$ scalar field variables
\begin{mathletters}
							\label{eqn-Pi+Q-defn}
\begin{eqnarray}
P_i
	& = &	\del_i \phi
									\\
Q
	& = &	(\partial_t \phi - \beta^i \del_i \phi) / \alpha
\end{eqnarray}
\end{mathletters}
so that
\begin{mathletters}
						\label{eqn-partial-phi=fn(P+Q)}
\begin{eqnarray}
\partial_t \phi
	& = &	\alpha Q + \beta^k P_k
									\\
\partial_i \phi
	& = &	P_i
\end{eqnarray}
\end{mathletters}
(These definitions~\eqref{eqn-Pi+Q-defn}
and~\eqref{eqn-partial-phi=fn(P+Q)} are similar, but not identical,
to those of \citesprefix\cite{Choptuik-PhD,Choptuik-1991-consistency,
Marsa-PhD,Marsa-Choptuik-1996-sssf}.)  A straightforward calculation
then gives the $3+1$ energy and momentum densities as
\begin{mathletters}
\begin{eqnarray}
4 \pi \rho
	& = &	\thalf (P_k P^k + Q^2)
									\\
4 \pi j_i
	& = &	- P_i Q
\end{eqnarray}
\end{mathletters}
and the spatial stress-energy tensor and its trace as
\begin{eqnarray}
4 \pi T\ij
	& = &	P_i P_j + \thalf g\ij (- P_k P^k + Q^2)
									\\
4 \pi T
	& = &	- \thalf P_k P^k + \tfrac{3}{2} Q^2
\end{eqnarray}
Corresponding to~\eqref{eqn-York-xform}, the York-transformation
scalings for $P_i$ and $Q$ are
\begin{mathletters}
						\label{eqn-York-xform-PQ}
\begin{eqnarray}
(P_i)\sub_\out
	& = &	\Psi^{-2} P_i
									\\
(Q)\sub_\out
	& = &	\Psi^{-4} Q
\end{eqnarray}
\end{mathletters}

We now assume that spacetime is spherically symmetric, with the
spatial coordinates $x^i = (r,\theta,\phi)$ having the usual polar
spherical topology.  However, we leave $r$ arbitrary, \ie{} we make
no assumption about the choice of (the radial component of) the
shift vector.  We take the $3+1$ field tensors to have the
coordinate components
\begin{eqnarray}
g\ij
	& \equiv &
		\diag
		\left[
		\begin{array}{c@{\quad}c@{\quad}c}
		A	& B	& B \sin^2 \theta	
		\end{array}
		\right]
									\\
K\ij
	& \equiv &
		\diag
		\left[
		\begin{array}{c@{\quad}c@{\quad}c}
		X	& Y	& Y \sin^2 \theta	
		\end{array}
		\right]
									\\
\beta^i
	& \equiv &
		\left[
		\begin{array}{c@{\quad}c@{\quad}c}
		\beta	& 0	& 0			
		\end{array}
		\right]
									\\
P_i
	& \equiv &
		\left[
		\begin{array}{c@{\quad}c@{\quad}c}
		P	& 0	& 0			
		\end{array}
		\right]
\end{eqnarray}
and for the York decomposition we take
\begin{equation}
\Omega^i
	\equiv	\left[
		\begin{array}{c@{\quad}c@{\quad}c}
		\Omega	& 0	& 0			
		\end{array}
		\right]
\end{equation}

Notice that we do {\em not\/} factor out either a conformal factor
or any $r^2$ factors from the 3-metric components.
\footnote{
	 Although this somewhat simplifies the $3+1$
	 equations, it may degrade the accuracy of
	 our computational scheme at large $r$.  In
	 particular, in hindsight not factoring out
	 $r^2$ factors may have been an unwise design
	 choice.  It might be interesting to try a
	 side-by-side accuracy comparison between
	 otherwise-identical factored and unfactored
	 schemes, but we haven't investigated this.
	 }

With these assumptions, it's then straightforward, though
tedious, to express all the other $3+1$ variables in terms of the
\defn{state variables} $A$, $B$, $X$, $Y$, $P$, and $Q$, and their
spatial (1st and 2nd) derivatives.  For reference, in
appendix~\ref{app-sssf-eqns} we tabulate the resulting equations
for all the $3+1$ variables involved in our initial data algorithm.
%
%
\section{Numerical Methods}
\label{sect-numerical-methods}

We use standard finite differencing techniques to numerically solve
the York updating equation~\eqref{eqn-York-update}.  We describe
our numerical methods in detail in a following paper
(\citeprefix\cite{Thornburg-1998-sssf-evolution}); we merely
summarize them here.

For the initial data equations, we use the usual centered 5~point
4th~order finite difference molecules in the grid interior, and
off-centered 5~(6)~point 4th~order molecules for 1st~(2nd)~derivatives
near the grid boundaries.  We use the same finite difference molecules
in all contexts in the equations, including ``left hand side'',
``right hand side'', interior equations, and boundary conditions.

After finite differencing the elliptic PDE~\eqref{eqn-York-update},
we row scale the resulting linear system to improve its numerical
conditioning, then $\sf LU$ decompose and solve it via \code{LINPACK}
band matrix routines (\citeprefix\cite{LINPACK-book}).  We use IEEE
double (64~bit) precision for all floating point computations.

We use a smoothly nonuniform grid, chosen to be uniform in the
\defn{warped} radial coordinate
\begin{equation}
\wr(r) = \frac{r_0}{a} \left( 1 - \frac{1}{r/r_0} \right)
	 + \frac{r_0}{b} \log \left( \frac{r}{r_0} \right)
	 + \frac{r_0}{c} \left( \frac{r}{r_0} - 1 \right)
						\label{eqn-mixed-210-wr(r)}
\end{equation}
where $r_0$, $a$, $b$, and $c$ are suitably chosen parameters.
We don't use any sort of staggered grid.  Note that we use $\wr$
only for finite differencing -- all tensor components are still
taken with respect to the $(r,\theta,\phi)$ coordinate basis.

By convention, we always place the inner grid boundary at $\wr = 0$,
\ie{} $r = r_0 \equiv r_\min$.  All our results in this paper use
the parameters $r_0 = 1.5$, $a = \infty$ (effectively omitting the
first term in~\eqref{eqn-mixed-210-wr(r)}), $b = 5$, and $c = 100$.
As shown in figure~\ref{fig-mixed-210-coord}, for these parameters
$\wr$ qualitatively resembles a logarithmic radial coordinate in the
inner part of the grid, and a uniform radial coordinate in the outer
part of the grid.
%
%
\section{Sample Results}
\label{sect-sample-results}

As shown in table~\ref{tab-test-slice-pars}, we have computed a
number of test initial data slices, using various combinations of
numerical parameters (grid resolutions and outer boundary positions)
and initial perturbations.  Most of our discussion of these slices
is independent of their numerical parameters, so we usually refer
generically to the families of slices sharing common initial
perturbations, and plot only the highest-resolution results within
each family.

In describing our results, we use the subscript
``$_\init$'' to refer to the initial slice
	(step~\ref{algorithm-step-analytical-BH-slice}
	in our overall initial data algorithm of
	section~\ref{sect-initial-data-algorithm+ansatze}),
``$_\perturb$'' to refer to the perturbed slice
	(the result of step~\ref{algorithm-step-perturbation}),
``$_\York$'' to refer to the result of the York projection operator
	(step~\ref{algorithm-step-York-projection}),
and ``$_\final$'' to refer to the final result after the
numerical coordinate transformation back to an areal radial coordinate
	(step~\ref{algorithm-step-coord-xform}).

All our results presented here use an Eddington-Finkelstein slice of a
unit-mass Schwarzschild spacetime for the initial slice
	(step~\ref{algorithm-step-analytical-BH-slice}),
and include a final numerical 3-coordinate transformation to make
the radial coordinate areal
	(step~\ref{algorithm-step-coord-xform}).
%
%
\subsection{Scalar Field Phenomenology}
\label{sect-sample-results/scalar-field-phenom}

As a first example of our initial data solver, consider the pqw5
slice, defined by the perturbation
(step~\ref{algorithm-step-perturbation} in our overall initial
data algorithm of section~\ref{sect-initial-data-algorithm+ansatze})
\begin{equation}
P \to P + 0.02 \times \Gaussian(r_\init{=}20, \sigma{=}5)
						\label{eqn-pqw5-perturbation}
\end{equation}
($Q$ remains identically zero.)

Figure~\ref{fig-200.pqw5-h+Psi+Omega} shows the York-decomposition
variables $\Psi$ and $\Omega$ for this slice.  Notice that because
the York equation~\eqref{eqn-York} is elliptic, the localized
perturbation~\eqref{eqn-pqw5-perturbation} results in the York
projection operator being nontrivial ($\Psi \neq 1$, $\Omega \neq 0$)
throughout almost the entire slice.  However, for the relatively
small perturbation~\eqref{eqn-pqw5-perturbation}, the York projection
and the numerical coordinate transformation back to an areal radial
coordinate (the latter given here by $r_\final / r_\York = \Psi^2$)
both differ from the identity by only a few percent.

Figure~\ref{fig-200.pqw5-h+ABXY.ratio} shows the relative
3-metric and extrinsic curvature components for this slice,
and figure~\ref{fig-200.pqw5.h+K+R+RR.ratio} shows the resulting
relative 3-invariants $K$, $R$, and $R_{ij} R^{ij}$.  Due to the
nonlocality of the York projection operator, all the geometric
field variables deviate significantly from their initial
(Schwarzschild\,/\,Eddington-Finkelstein) values throughout the
slice.  Notice also that there's an apparent horizon just inside
$r = 2$, \ie{} the slice does (still) contain a black hole.

Figure~\ref{fig-200.pqw5-h+scalar-field+mass} shows this slice's
scalar field and mass distributions.  As can be seen, the slice
contains a shell of scalar field surrounding the black hole.  The
scalar field shell has a roughly Gaussian profile, which for the
radial density $4 \pi B \rho$ is centered at $r \approx 21.8$, with
thickness $\sigma \approx 3.5$.  Despite the relatively small
perturbation~\eqref{eqn-pqw5-perturbation}, the scalar field shell
is fairly massive, approximately $0.66$~times the black hole mass
(approximately $40\%$ of the slice's total mass).  [Although it's
not apparent here, time-evolving this initial data
(\citeprefix\cite{Thornburg-1998-sssf-evolution}) shows that the
shell is actually the superposition of two momentarily coincident
shells, one ingoing and one outgoing.  However, contrary to what
one might expect, the two shells have quite different masses.]
%
%
\subsection{Spacetime and Slicing Phenomenology}
\label{sect-sample-results/spacetime+slicing-phenom}

From figure~\ref{fig-200.pqw5-h+scalar-field+mass} it's clear that
away from the scalar field shell the pqw5 slice is nearly vacuum
(all our scalar field diagnostics are very small), so by Birkhoff's
theorem it must be nearly Schwarzschild there.  More precisely,
inside the scalar field shell the pqw5 slice must be nearly
(isometric to) some slice of a Schwarzschild spacetime with mass
$m_< \equiv m_\bh \approx 0.976$, while outside the scalar field
shell it must be nearly (isometric to) some slice of a Schwarzschild
spacetime with mass $m_> \equiv m_\total \approx 1.617$.

In support of this,
figure~\ref{fig-200.pqw5-h+four-Riem-Riem.mass-ratio} shows the
mass-relative 4-Riemann curvature invariant $R_{abcd} R^{abcd}$
for the pqw5 slice.  As expected, away from the scalar field shell
$R_{abcd} R^{abcd}$ is almost exactly equal to its value for a
Schwarzschild spacetime of matching mass ($m_<$ inside the shell,
$m_>$ outside).

Figure~\ref{fig-200.pqw5.h+K+R+RR.mass-ratio} shows the mass-relative
3-invariants $K$, $R$, and $R_{ij} R^{ij}$ for the pqw5 slice.  The
mass-relative ratios all deviate substantially from unity even in
the near-vacuum regions.  From this we conclude that although the
pqw5 slice is nearly (isometric to) {\em some\/} slice of a
matching-mass Schwarzschild spacetime in each of its near-vacuum
regions, neither of these slices is our canonical Eddington-Finkelstein
one.

It might be interesting to explicitly compute the embedding of these
regions of the slice in their respective Schwarzschild spacetimes,
\ie{} explicitly compute (say) each region's Eddington-Finkelstein
or Schwarzschild time coordinate as a function of spatial position
(\eg{} areal radius $r$).  The two time coordinates would each only
be defined up to an arbitrary additive constant, and (being referred
to different Schwarzschild spacetimes) they wouldn't be directly
comparable, but their intra-region variation might still be
interesting.  However, we haven't as yet made such a study.

From figures~\ref{fig-200.pqw5.h+K+R+RR.ratio}
and~\ref{fig-200.pqw5.h+K+R+RR.mass-ratio}, it appears that
at large $r$, $K/K_\Schw \to 1$ and $R/R_\Schw \to 1$, while
$J \equiv R_{ij} R^{ij}$ displays the different behavior
$J / J_{\Schw(m(r))} \to 1$ and hence (\cf{}~\eqref{eqn-Schw/EF-RR})
$J / J_\Schw \to (m_\total/m_\init)^2$.  Examining other
slices computed with a larger outer boundary radius, \eg{} the
100o30.pqw5 and 200o30.pqw5 slices (\cf{}~figure~
\ref{fig-100o30+200o30.pqw5-outer-curvature-ratio+mass-ratio})
confirms these asymptotic behaviors.  A simple analytical argument
also confirms this behavior for $K$:  The
transformation~\eqref{eqn-York-xform} in the York projection
operator preserves $K$ at any given event, and since $\Psi \to 1$
at large $r$, the transformation back to an areal radial coordinate
preserves $K$'s leading order behavior there.  Thus our initial
data slices must have asymptotically the same $K$ as the initial
Schwarzschild\,/\,Eddington-Finkelstein slice, \ie{}
$K / K_{\Schw(m_\init)} \to 1$ at large $r$.

Since our different 3-invariant diagnostics have different
asymptotic behaviors relative to our canonical Eddington-Finkelstein
slice, it's clear that our slice isn't (even) asymptotically
Eddington-Finkelstein.  This asymptotic ``warping'' of the slice
may prove a complication for $3+1$ time evolution calculations.  It
might be worth investigating whether modifications to our initial
data algorithm (\eg{} a different type of perturbation, and/or
different outer boundary conditions) might yield slices with
nicer asymptotic properties, but we haven't as yet done this.
%
%
\subsection{Accuracy}
\label{sect-sample-results/accuracy}

The main error sources in a $3+1$ code such as ours are due to
the finite grid resolution (finite differencing truncation errors),
the finite outer boundary radius, and floating point roundoff.
As discussed in appendix~\ref{app-convergence-tests}, we expect
the first two of these to vary in predictable ways with the
corresponding numerical parameters (grid resolutions and outer
boundary positions), and we can quantitatively measure these errors
by comparing computations of the same physical system using different
numerical parameters.  In the context of a finite differencing
code such as ours, \ie{} one solving elliptic PDEs such as the
York equation~\eqref{eqn-York}, floating point roundoff errors
are easily distinguished from other error sources by their
non-smoothness (large and quasi-random variations from
one grid point to the next).

As an example of our computational scheme's overall accuracy
level, figure~\ref{fig-100+200.pqw5.C+C-r.Schw+perturb+final}
shows the magnitude of the numerically computed energy and momentum
constraints at various points in the initial data computation for
the 100.pqw5 and 200.pqw5 slices.  Consider first part~(a) of the
figure, which shows the magnitude $|C|$ of the energy constraint,
and for the moment focus only on the 100.pqw5 curves:  $|C_\Schw|$
shows the basic ``background'' accuracy level of our finite
differencing, tailing off into floating point roundoff noise for
$r \gtsim 40$.  In contrast, $|C_\perturb|$ shows the effects of
the perturbation~\eqref{eqn-pqw5-perturbation}: $|C_\perturb|$ is
large (${\gg}\, |C_\Schw|$) wherever the perturbation is significant.
Finally, $|C_\final|$ shows the effects of the York projection
operator (and the numerical coordinate transformation back to
an areal radial coordinate), with $|C_\final| \ll |C_\perturb|$
wherever the perturbation is significant, and more generally
$|C_\final|$ small (comparable in magnitude to $|C_\Schw|$)
everywhere in the slice.  In other words, we see that the York
projection operator does indeed act to retore the field variables
back into the constraint hypersurface.

To assess our computational scheme's finite differencing errors,
figure~\hbox{\ref{fig-100+200.pqw5.C+C-r.Schw+perturb+final}(a)} also
shows $|C_\Schw|$, $|C_\perturb|$, and $|C_\final|$ for the
200.pqw5 slice, which (\cf{}~table~\ref{tab-test-slice-pars})
is computed using twice the grid resolution (half the grid spacing)
as the 100.pqw5 slice, but is otherwise identical.  Notice that for
$|C_\Schw|$ and $|C_\final|$, the 100.pqw5 and 200.pqw5 curves are
both very similar in shape, the 200.pqw5 curve being offset downwards
(smaller $|C|$) by a factor of~$16$ (compare with the scale bar),
until both tail off into roundoff noise at large $r$.  As discussed
in appendix~\ref{app-convergence-tests}, this constitutes strong
evidence that these $|C|$ values are indeed dominated by 4th~order
finite differencing truncation errors at small and moderate $r$,
and floating point roundoff errors at large $r$.
\footnote{
\label{footnote-boundary-FD-effects-in-C-and-C-r}
	 Due to boundary finite differencing effects
	 (\cf{}~appendix~\ref{app-convergence-tests}),
	 $C$ and $C^r$ are both roughly a factor of
	 $10$ larger at the grid boundaries than at
	 nearby non-boundary grid points.  However,
	 even the boundary $C$ and $C^r$ values still
	 show 4th~order convergence.
	 }
{}  Notice also that the overall level of the $|C|$ values, \ie{}
of the deviation of the energy constraint from satisfaction, is
very low, with $|C_\final| \ltsim 10^{-8}$ ($10^{-9}$) for the
100.pqw5 (200.pqw5) slice.

Figure~\hbox{\ref{fig-100+200.pqw5.C+C-r.Schw+perturb+final}(b)} shows
the magnitude of the numerically computed momentum constraint $C^r$\,
\footnote{
	 In spherical symmetry $C^r$ is the only
	 nonzero component of the momentum constraint
	 vector $C^i$, \cf{}~\eqref{eqn-sssf/C-i-vector}.
	 }
{} at the same points in the initial data computation for these
slices.  (Since the perturbation~\eqref{eqn-pqw5-perturbation}
doesn't change $C^r$, $|C^r_\perturb| = |C^r_\Schw|$, so these are
plotted as the same curve.)  From the figure it's clear that the
overall level of the momentum constraints is ${\sim}\, 3$~orders
of magnitude smaller than that of the energy constraint.  Also,
$|C^r_\final|$ is small (comparable in magnitude to $|C^r_\Schw|$)
everywhere in the slice, and both show 4th~order convergence in the
same manner as $|C_\Schw|$ and $|C_\final|$.

As an example of a more quantitative test of the finite differencing
errors, figure~\ref{fig-100+200.pqw5.C.conv} shows a scatterplot of
the 200.pqw5 versus the 100.pqw5 $C_\final$ values.  As discussed
in appendix~\ref{app-convergence-tests}, for 4th~order convergence
all the points (except for outliers from the grid boundaries) should
fall on a line through the origin with slope~$\tfrac{1}{16}$.  As
can be seen, except for the grid-boundary outliers, all the points
do indeed fall very closely on this line.  We emphasize here that
(\cf{}~appendix~\ref{app-convergence-tests}) the plotted line is
{\em not\/} a fit to the data, but rather an a~priori prediction
with {\em no\/} adjustable parameters, making this a very stringent
test.

Errors due to the finite outer boundary radius $r_\max$ are
more problematical: these enter through the outer boundary
conditions~\eqref{eqn-Psi-outer-BC-scalar-Robin}
and~\eqref{eqn-Omega-outer-BC-scalar-Robin}, which
essentially fix the asymptotic flatness of the initial data
slice.  Without a clear-cut measurement of the slice's
embedding in the far-field Schwarzschild region
(\cf{}~section~\ref{sect-sample-results/spacetime+slicing-phenom}),
asymptotic-flatness errors are difficult to measure.  Moreover,
it's not immediately obvious just how (\ie{} with what power of
$r_\max$) these errors should scale with $r_\max$.

However, to give an indication of the typical magnitudes of
these errors, the relative changes $\delta K/K$, $\delta R/R$,
$\delta J/J$, and $\delta I/I$ (where $J \equiv R_{ij} R^{ij}$
and $I \equiv R_{abcd} R^{abcd}$) between the same positions in
(common to) the pqw5 slices with $w_\max = 4$ ($r_\max \approx 248$)
and $w_\max = 10$ ($r_\max \approx 813$), are all on the order
of $10^{-5}$, while the relative changes $\delta m/m$ in the
computed black hole and total spacetime masses are
${\sim}\, 3{\times}10^{-6}$.  The corresponding changes between
the slices with $w_\max = 10$ ($r_\max \approx 813$) and
$w_\max = 30$ ($r_\max \approx 2\,780$), are all about a factor
of $10$ smaller, and in general the outer boundary errors appear
to scale approximately as ${\sim}\, 1 / r_\max{}^2$.  These are
all {\em continuum\/} effects: with one exception discussed below,
the differences between different-resolution slices with the
same $r_\max$, are all much smaller than the differences between
slices with different $r_\max$.

The one significant non-continuum effect apparent in these
slices is the presence of significant floating-point roundoff noise
in our curvature diagnostics, particularly in $R$, at large $r$.
As an example of this,
figure~\ref{fig-100o30+200o30.pqw5-outer-curvature-ratio+mass-ratio}
shows the relative and mass-relative values of various embedding
and curvature diagnostics for the outer parts of the 100o30.pqw5
and 200o30.pqw5 slices.  While the $K$, $R_{ij} R^{ij}$, and
and $R_{abcd} R^{abcd}$ values show no visible noise, the $R$
values show serious noise, with the relative noise amplitude
$\delta R/R_\Schw$ roughly 4~times larger in the 200o30.pqw5 slice,
and in both slices growing rapidly with $r$ to ${\sim}\, 5\%$ ($20\%$)
at the outer boundary ($r_\max \approx 2\,780$) of the 100o30.pqw5
(200o30.pqw5) slice.  The rapid growth of the relative noise
amplitude with $r$ is actually entirely due to the rapid
(${\sim}\, 1/r^4$) falloff in $R_\Schw$: within each slice the
actual absolute noise level in $R$ is essentially constant for
the range of $r$ in the figure.

Comparing the computations (appendix~\ref{app-sssf-eqns}) of the
different diagnostics plotted in
figure~\ref{fig-100o30+200o30.pqw5-outer-curvature-ratio+mass-ratio},
we see that $K$ is computed from the 3-metric and extrinsic curvature
components by algebraic operations only, while $R$, $R_{ij} R^{ij}$,
and $R_{abcd} R^{abcd}$ all depend on (2nd)~numerical derivatives
of the 3-metric components.  Numerical differentiation is well known
as a noise-amplifying process, and will inherently amplify normal
low-level roundoff noise in the 3-metric components.  To test whether
this process can account for the observed noise in $R$,
figure~\ref{fig-100o30+200o30.pqw5-outer-curvature-ratio+mass-ratio}
also shows the relative value of $R_\fit \pm \varepsilon/(\Delta \wr)^2$,
where $R_\fit$ is an empirical least-squares fit to the $R$ values,
and the \,$\pm \varepsilon/(\Delta \wr)^2$\, term models the effects
on $R$ of $O(\varepsilon)$ roundoff errors in the 3-metric components
being 2nd-differentiated, with $\varepsilon = 6{\times}10^{-19}$ an
``eyeball-fitted'' constant parameter, and $\Delta \wr$ the grid
spacing in our $\wr$ nonuniform gridding coordinate
(\cf{}~section~\ref{sect-numerical-methods}).  From the figure it's
clear that this model does in fact nicely account for the $r$-~and
$\Delta \wr$-dependence of $R$'s roundoff noise.

Although this explains the noise in $R$, it raises the further
question of why $R_{ij} R^{ij}$ and $R_{abcd} R^{abcd}$ don't show
similar effects.  (A careful analysis does show some roundoff noise
in $R_{ij} R^{ij}$ and $R_{abcd} R^{abcd}$, but only at levels many
orders of magnitude below that in $R$.)  We're not certain, but we
suspect the explanation lies in the detailed form of the respective
equations, with the offending 2nd~derivative terms being (presumably)
relatively less important in the computations of $R_{ij} R^{ij}$
and $R_{abcd} R^{abcd}$ than in $R$.  In any case, though, the
absolute magnitude of even $R$'s roundoff noise is still very small,
$\pm \varepsilon/(\Delta \wr)^2 \approx 6{\times}10^{-15}$
($2{\times}10^{-14}$) for the 100o30.pqw5 (200o30.pqw5) slice.
Although $R$ appears as a coefficient in the York
equation~\eqref{eqn-York}, the offending term is linear (so $R$'s
noise should mainly average out), and the noise doesn't appear to
be a serious error source in the solutions $\Psi$ and $\Omega^i$,
or in the computed 3-metric and extrinsic curvature components.
%
%
\subsection{Inner Boundary Condition}
\label{sect-sample-results/inner-BCs}

Thus far we have considered only the
perturbation~\eqref{eqn-pqw5-perturbation}, which is relatively small
at the inner boundary ($\delta P_\inner / \delta P_\max \approx 10^{-3}$).
As discussed in section~\ref{sect-initial-data-algorithm+ansatze},
our motivation for the inner boundary condition~\eqref{eqn-York-inner-BC}
is actually predicated on the perturbation having negligible effects
there.

It's thus of interest to determine whether and how well our initial
data algorithm works for perturbations which don't satisfy this
restriction, \ie{} for perturbations which have non-negligible
effects at the inner boundary.  As an example of such a perturbation,
consider the pqw5i slice.  As shown in table~\ref{tab-test-slice-pars},
this uses the perturbation
\begin{equation}
P \to P + 0.02 \times \Gaussian(r_\init{=}10, \sigma{=}5)
						\label{eqn-pqw5i-perturbation}
\end{equation}
instead of~\eqref{eqn-pqw5-perturbation}.  The
perturbation~\eqref{eqn-pqw5i-perturbation} has the same amplitude
and width as~\eqref{eqn-pqw5-perturbation}, but is centered at
$r_\init = 10$ instead of $r_\init = 20$, so it's fairly large at
the inner boundary ($\delta P_\inner / \delta P_\max \approx 0.24$).

Figure~\ref{fig-100+200.pqw5i.C.Schw+perturb+final} shows $|C_\Schw|$,
$|C_\perturb|$, and $|C_\final|$ for the pqw5i slice, analogously to
figure~\hbox{\ref{fig-100+200.pqw5.C+C-r.Schw+perturb+final}(a)} for
the pqw5 slice.  It's clear that despite the pqw5i slice's substantial
perturbation at the inner boundary, its initial data computation
remains fully accurate, with $|C_\final| \ll |C_\perturb|$ wherever
the perturbation is significant, and $|C_\final|$ small (comparable
in magnitude to $|C_\Schw|$) everywhere in the slice, including near
to and at the inner boundary.

Moreover, comparing the $|C_\Schw|$ and $|C_\final|$ curves in
figure~\ref{fig-100+200.pqw5i.C.Schw+perturb+final}, we see that
the 100.pqw5i and 200.pqw5i curves are again both very similar
in shape, the 200.pqw5i curve being offset downwards (smaller $|C|$)
by a factor of $16$, until both tail off into roundoff noise at
large $r$.  Just as for the pqw5 slice (\cf{}~our discussion of
figure~\ref{fig-100+200.pqw5.C+C-r.Schw+perturb+final}
in section~\ref{sect-sample-results/accuracy}), this implies that
these errors (the nonzero values of $C$) are dominated by (the
expected) 4th~order finite differencing truncation errors at small
and moderate $r$, and floating point roundoff errors at large $r$.
\footnote{
	 The pqw5i $|C|$ values in
	 figure~\ref{fig-100+200.pqw5i.C.Schw+perturb+final}
	 also show similar boundary finite differencing
	 anomalies (\cf{}~appendix~\ref{app-convergence-tests}),
	 to the pqw5 $|C|$ values in
	 figure~\ref{fig-100+200.pqw5.C+C-r.Schw+perturb+final}
	 (\cf{}~footnote~\ref{footnote-boundary-FD-effects-in-C-and-C-r}).
	 }
{}  The same is also true of $C^r$ for these slices (details of the
analysis omitted for brevity).

These pqw5i results demonstrate that despite our inner boundary
condition~\eqref{eqn-York-inner-BC} being motivated by -- and only
for -- the special case where the perturbation's effects are negligible
at the inner boundary, we actually obtain fully accurate initial data
even for perturbations which are large there.
%
%
\subsection{Other Perturbations}
\label{sect-sample-results/other-perturbations}

As shown in table~\ref{tab-test-slice-pars}, we have also
experimented with several alternative variants of the
perturbation~\eqref{eqn-pqw5-perturbation} for
step~\ref{algorithm-step-perturbation} in our algorithm
summary of section~\ref{sect-initial-data-algorithm+ansatze}.

Figure~\ref{fig-200.pqw5b+pqw5c-h+scalar-field+mass} shows the
effect of increasing the perturbation amplitude from $0.02$ (the
pqw5 slice discussed above) to $0.05$ (the pqw5b slice) or $0.1$
(the pqw5c slice).  With one exception disussed below, all three
slices are qualitatively fairly similar.  In going from the pqw5
slice to the pqw5b and pqw5c slices, the main detailed differences
are that as a result of the larger perturbations, the scalar field
shells become much more massive (from approximately $0.66$ to
$3.9$~and $17$~times the black hole mass, or equivalently from
approximately $40\%$ to $80\%$~and $94\%$~of the total spacetime
mass, respectively), and are also somewhat larger and thicker.
As well, the deviations (not shown here) of all the geometric
field variables from their Schwarzschild\,/\,Eddington-Finkelstein
values become much larger.  Despite these stronger nonlinear
effects, we experienced no difficulties with poor convergence
of the Newton-Kantorovich iteration for these models.
\footnote{
	 In retrospect this isn't too surprising,
	 as even for the pqw5c slice the peak
	 values of $\Psi - 1$ and $\Omega$ are
	 still only~${\approx}\, 0.22$ and~$0.16$
	 respectively, so the nonlinearities are
	 still fairly weak in an absolute sense.
	 }

The pqw5c slice shows a new effect: although the mass function still
shows a significant mass contained within the numerical grid's inner
boundary (at $r = 1.5$), here there's no longer any apparent horizon
within the numerical grid.  The pqw5c slice's ``black hole'' mass
shown in table~\ref{tab-test-slice-pars} thus no longer represents
the mass of an actual black hole, but rather merely of the region
within the inner boundary.  Since $r \approx 2.6m > 2m$ on the inner
boundary this region may, but need not necessarily, contain a black
hole.  (Even with the constraints imposed, the continuation of the
field variables into the origin is obviously non-unique.)

Since the inner grid boundary for this slice doesn't lie within an
apparent horizon, this numerical data is unsuitable for a
black-hole--excluded time evolution.  However, viewed purely as
initial data, it remains fully accurate, \ie{} the constraints are
small everywhere in the numerical grid.

Figure~\ref{fig-400.pqw1-h+scalar-field+mass} shows the effects
of using a narrow (width-$1$) but high-amplitude ($0.1$) perturbation
(the pqw1 slice).  This slice qualitatively resembles the pqw5b
slice: it contain a black hole surrounded by a moderately massive
scalar field shell (approximately $3.0$~times the black hole mass,
or equivalently approximately $75\%$ of the total spacetime mass).
However, corresponding to the narrow perturbation, here the scalar
field shell is very thin, with a fractional width ($\sigma/r_\ccenter$
for a Gaussian fit to the radial scalar field density $4 \pi B \rho$)
of only $0.033$.

The pw5+qw3 slice shows the effect of applying perturbations to both
$P$ and $Q$, instead of $P$ alone as in the previous slices.  This
slice is qualitatively fairly similar to the pqw5b slice, except that
since both $P$ and $Q$ are nonzero, $j^r$ is nonzero as well as $\rho$.

Having $j^r$ nonzero makes this slice particularly useful for testing
the equality of the Misner-Sharp and integrated--scalar-field mass
functions $m_\MS$ and $m_\mu$ (\cf{}~section~\ref{sect-diagnostics}),
since now both terms in the definition~\eqref{eqn-mu} of $\mu$ contribute,
instead of just the $\rho$ term as in all the previous slices.
Figure~\ref{fig-100+200.pw5+qw3.h+MS-mass.relerr} shows the relative
deviation of $m_\MS$ from $m_\mu$ for this slice, for both grid
resolutions.  Both deviations are very small across their slices,
with $m_\MS = m_\mu$ to a relative accuracy of ${\ltsim}\, 10^{-5}$
($10^{-6}$) for the 100.pq5+qw3 (200.pq5+qw3) slice.  Moreover, a
quantitative convergence test (\cf{}~appendix~\ref{app-convergence-tests}
and section~\ref{sect-sample-results/accuracy}; details omitted
for brevity) shows that these deviations are once again dominated
by (the expected) 4th~order finite differencing truncation errors.

Instead of perturbing the matter variables, we can perturb the
initial slice's geometry (3-metric and/or extrinsic curvature
components).
\footnote{
	 Or we could perturb both matter and
	 geometry.
	 }
{}  Since our initial data \Ansatze{} always take the initial
analytical black hole slice to be vacuum, and the York projection
operator~\eqref{eqn-York-xform} and~\eqref{eqn-York-xform-PQ}
obviously preserves this, the resulting slice will therefore
also always be vacuum.  We have previously shown
(\citeprefix\cite{Thornburg-PhD}, appendix~5) that in the
non--spherically-symmetric case, suitable geometry perturbations
yield (vacuum) dynamic initial slices containing black holes
surrounded by gravitational radiation.

However, in spherical symmetry any vacuum slice is necessarily some
slice in (a) Schwarzschild spacetime.  Although a ``warped Schwarzschild''
slice of this type has no true dynamics, it can still be useful as
a numerical-relativity test-bed problem:  None of the test problems
suggested by \citeprefix\cite{CSEHT-1986-in-Centrella} are
applicable to the spherically symmetric scalar field system, but
\citeprefix\cite{York-1989-in-Frontiers} has suggested testing the
numerical evolution of ``hand-warped'' slices in Minkowski spacetime.
In the spirit of this latter suggestion, our ``warped Schwarzschild''
slices are of interest as alternative test cases.  (As will be seen,
they also offer a particularly good environment for testing an
initial data problem algorithm and computer code.)

To assess the accuracy of these slices, we use $m_\MS$ and
$R_{abcd} R^{abcd}$ as diagnostics:  $m_\MS$ should be constant across
each slice, and $R_{abcd} R^{abcd}$ should be everywhere the same as
in a matching-mass Schwarzschild spacetime (\ie{} the mass-relative
$I \equiv R_{abcd} R^{abcd}$ should be unity across the slice).

For example, consider the daw5c slice, which uses a perturbation
applied to $A \equiv g_{rr}$,
\begin{equation}
A \to A + 0.1 \times \Gaussian(r_\init{=}20, \sigma{=}5)
						\label{eqn-daw5-perturbation}
\end{equation}
This amplitude gives roughly the same warping of the slice
(\eg{} deviation of $K/K_\Schw$ from unity) as for the pqw5 slice.

Figure~\ref{fig-100+200.daw5c.h+MS-mass+four-Riem-Riem.relerr} shows
the relative deviations from unity (errors) of $m_\MS/m_\total$ and
$I/I_{\Schw(m_\total)}$ (\ie{} the relative errors in $m_\MS$ and $I$)
across the daw5c slice, computed with two different grid resolutions,
where $m_\total$ is the total mass of the slice (since the slice is
vacuum, this is equal to the Misner-Sharp mass within the inner grid
boundary).  Both diagnostics show only very small deviations across
the slices, with $m = m_\total$ to an relative accuracy of
${\ltsim}\, 10^{-5}$ ($5{\times}10^{-7}$) for the 100.daw5 (200.daw5)
slice, and $I = I_{\Schw(m_\total)}$ to a relative accuracy of
${\ltsim}\, 3{\times}10^{-4}$ ($5{\times}10^{-5}$) respectively.
Moreover, a quantitative convergence test
(\cf{}~appendix~\ref{app-convergence-tests} and
section~\ref{sect-sample-results/accuracy}; details again omitted
for brevity) shows that these errors are once again dominated
by (the expected) 4th~order finite differencing truncation errors.

Finally, note that the computation of $I \equiv R_{abcd} R^{abcd}$
via~\eqref{eqn-four-Riem-Riem}
and~\hbox{\eqref{eqn-sssf/four-Riem-Riem/first-subexpr}
--\eqref{eqn-sssf/four-Riem-Riem/last-subexpr}} depends nontrivially
on all the state variables (3-metric, extrinsic curvature, and scalar
field components) in our computational scheme.  The observation that
$I = I_{\Schw(m_\total)}$ to high accuracy for the vacuum slices, and
more generally that $I = I_{\Schw(m(r))}$ for the vacuum regions of all
our slices, thus gives a strong test of the overall correctness of a
large part (essentially the geometry terms in all the equations) of
our code, including freedom from most errors in either deriving our
continuum equations, or programming the numerical computations.
%
%
\section{Conclusions and Directions for Further Research}
\label{sect-conclusions}

Like many other researchers, we find York's conformal-decomposition
technique to offer an elegant, reliable, and fairly efficient means
of constructing $3+1$ initial data.  In this paper we consider the
general case of the York decomposition, where (as is required by
most black-hole--exclusion slicings) $K$ is nonzero and spatially
variable throughout most or all of the slices, and hence the full
nonlinear 4-vector York equations must be solved numerically.

The main focus of this paper is on the combination of this numerical
solution, with a set of \Ansatze{} to provide the York decomposition's
inputs and inner boundary conditions:  We begin with a known black
hole initial data slice (\Ansatz{}), apply an arbitrary (generally
constraint-violating) perturbation (\Ansatz{}) to it, then use the
York decomposition to project the perturbed field variables back into
the constraint hypersurface, using a further \Ansatz{} to supply inner
boundary conditions for the York equations.  (Our algorithm also
incorporates a final numerical coordinate transformation to (\eg{})
restore an areal radial coordinate, but this is only for convenience.)

In comparison to other initial data algorithms, the key advantage
of this algorithm is its flexibility:  It's easily applicable to
almost any slicing, and so permits the slicing to be chosen for
convenience in (say) a black-hole--excluded time evolution, rather
than being restricted by the initial data solver.  This algorithm
is also equally applicable in spherical symmetry, axisymmetry, or
in fully 3-dimensional spacetimes, and in both vacuum and non-vacuum
spacetimes.

Collectively, this algorithm's \Ansatze{} provide a great deal of
flexibility in controlling the physical content of the resulting
initial data.  An interesting area for further research would be
in trying to better understand the relationship between this physical
content, and the choices made for the algorithm's various \Ansatze{}.
That is, at present we can't a~priori (\ie{} before solving the
York equations) predict very much about the physical content of
the resulting initial data obtained from a given set of \Ansatze{}
choices, so if some particular type of initial data is desired, some
trial and error may be needed to obtain the desired results.  A better
understanding of this would both be of practical benefit in more
easily constructing initial data with desired properties, and quite
likely prove informative in mapping the range and phenomenology
of possible initial data obtainable using our methods.

Turning now to the individual \Ansatze{}, the choice of the
initial known black hole slice seems fairly straightforward:
Given the black-hole--exclusion requirement that the slicing and
all the $3+1$ field variables be nonsingular near to and on the
horizon, the natural choice is an Eddington-Finkelstein slice of
Schwarzschild spacetime in spherical symmetry, or a Kerr slice
of Kerr spacetime in more general cases.
\footnote{
	 Another option, which we haven't explored,
	 would be to use an initial slice constructed
	 by some numerical method, either this one or
	 another.  Or one could use a non--black-hole
	 initial slice (\eg{} a slice in Minkowski
	 spacetime), and try to choose the remaining
	 \Ansatze{} so as to produce a ``geon''
	 (\cf{}~\citeprefix\cite{Abrahams-Evans-1992-BH-geon-formation}).
	 }

In contrast, the choices for our other \Ansatze{} (the perturbation
to apply to the initial slice, and the inner boundary conditions
for the York decomposition) are less clear-cut.  The choices presented
here (respectively adding a Gaussian to one or a few of the 3-metric,
extrinsic curvature, or matter field variable components; and Dirichlet
inner boundary conditions) are essentially the first ones we tried,
and as yet we have made only limited experiments with other choices.
The choices presented here seem to work adequately, in the sense of
yielding dynamic black hole spacetimes suitable for numerical evolutions
(\citeprefix\cite{Thornburg-PhD,Thornburg-1998-sssf-evolution}), but
clearly these choices are only a tiny subset of the possible parameter
space.  Both numerical and theoretical studies of a broader range
of choices would be useful here.

As well as a general discussion of the initial data problem algorithm
on generic slices, in this paper we present numerical results for
a specific model problem system, the spherically symmetric scalar
field.  We find that the resulting full nonlinear 4-vector York
equations can be solved robustly and with high accuracy using a
Newton-Kantorovich ``outer linearization'', followed by a standard
direct solution of the resulting sequence of ``inner'' linear
elliptic PDEs, using 4th~order finite differencing on a smoothly
nonuniform grid.  It seems likely that a wide range of other
numerical methods would also work well on this problem.

We haven't encountered any difficulties with the York equations
failing to have valid and physically reasonable solutions, even
with strong perturbations yielding very massive scalar field shells
(up to ${\sim}\, 17$~times the black hole's mass in our tests).
However, we also don't have any rigorous proofs that such difficulties
couldn't arise for different choices of the various \Ansatze{}.
Further research in this direction would be valuable.

As examples of our initial data solver, we discuss a number of
Eddington-Finkelstein--like initial data slices containing black
holes surrounded by scalar field shells, and also several vacuum
black hole slices with nontrivial slicings.  As described in
table~\ref{tab-test-slice-pars}, all of our test slices use (local)
Gaussian perturbations added to one or two of the scalar field,
3-metric, or extrinsic curvature components.  The slices vary in
their perturbations' positions, amplitudes, and widths, in which
field variable or variables is/are perturbed, and in the numerical
parameters (grid resolution and outer boundary position).

We use a number of diagnostics to analyze these slices, including
apparent horizon positions, the Misner-Sharp and integrated--scalar-field
mass functions $m_\MS$ and $m_\mu$ (respectively), the scalar field's
3-energy density $\rho$ and radial 3-energy density
$4 \pi g_{\theta\theta} \rho$, the extrinsic curvature scalar $K$,
the 3-Ricci scalar $R$ and quadratic curvature invariant $R_{ij} R^{ij}$,
the 4-Ricci scalar $\four\! R$, and the 4-Riemann quadratic curvature
invariant $R_{abcd} R^{abcd}$.

The focus of our analysis of the test slices is on exploring the
general phenomenology of the initial data slices, and on testing
and validating the accuracy of the initial data algorithm itself.
The computed slices' scalar field shells have masses ranging from as
low as $0.17$ to as high as $17$~times the black hole mass $m_\bh$,
\ie{} the scalar field shells comprise from as little as $14\%$
to as much as $94\%$ of the total slice masses.  The scalar field
shells have approximately Gaussian profiles with fractional widths
$\sigma / r_\ccenter$ (for the radial scalar field density
$4 \pi g_{\theta\theta} \rho$) ranging from $3.3\%$ to $14\%$, and
positions (radii) ranging from $13 m_\bh$ to $28 m_\bh$.

In all cases we find that the computed slices are very accurate:
For grids with resolutions of $\Delta r/r \approx 0.02$ ($0.01$)
near the perturbations, the numerically computed energy constraints
are ${\ltsim}\, 10^{-8}$ ($10^{-9}$) in magnitude; the numerically
computed momentum constraints are typically ${\sim}\, 3$~orders
of magnitude smaller.  The Misner-Sharp and integrated-scalar-field
mass functions agree to within relative accuracies of
${\ltsim}\, 10^{-6}$ ($10^{-7}$) respectively.  In tests of numerically
constructed vacuum slices with nontrivial slicings, where we would
expect the Misner-Sharp mass function to be constant and the 4-Riemann
quadratic curvature invariant $R_{abcd} R^{abcd}$ to be equal to its
Schwarzschild-spacetime value, we find these properties to hold to
within relative errors of ${\ltsim}\, 10^{-5}$ ($5{\times}10^{-7}$)
and ${\ltsim}\, 3{\times}10^{-4}$ ($5{\times}10^{-5}$) respectively.
Except for floating point roundoff noise (most prominently in the
3-Ricci scalar $R$), the errors we do see in these and other
diagnostics are all generally dominated by the expected
$O \bigl( (\Delta r)^4 \bigr)$ finite differencing truncation errors.

As well as our main focus on the initial data problem, in the
appendices of this paper we also present results in several related
peripheral areas, including a tabulation of some of the key $3+1$
field variables for an Eddington-Finkelstein slice of Schwarzschild
spacetime, the formulation of the Misner-Sharp mass function without
restrictions on the slicing or spatial coordinates, numerical
3-coordinate transformations, and tests of the convergence of finite
differencing computations to the continuum limit.  We also review
some little-known, but surprising, numerical analysis results on
the non-smoothness of the errors incurred when interpolating data
from one grid to another by the usual moving-local-interpolation
schemes, and discuss these results' import for numerical coordinate
transformations and horizon finding.
%
%
\section*{Acknowledgments}
\label{sect-acknowledgments}

We thank Niall \OMurchadha{} for invaluable assistance with the
definition of $\mu$, and for suggesting the argument given in
appendix~\ref{app-sssf-eqns} for why apparent horizons in spherical
symmetry always have $r_\areal = 2m$.
We thank M.~Huq for useful discussions on the non-smoothness of
interpolation errors.
We thank W.~G.~Unruh, R.~Parachoniak,
the University of British Columbia Physics Department,
R.~A.~Matzner,
and the University of Texas at Austin Center for Relativity
for their hospitality and the use of their research facilities
at various times during the course of this work.
We thank J.~Wolfgang, R.~Parachoniak, P.~Luckham, and J.~Thorn
for major assistance with setting up and maintaining computer facilities,
and G.~Rodgers and J.~Thorn for financial support.
%
%
\appendix
%
%
\section{Schwarzschild Spacetime in Eddington-Finkelstein Coordinates}
\label{app-Schw-EF}

Although the Eddington-Finkelstein coordinates $(t,r,\theta,\phi)$ for
Schwarzschild spacetime are well known (\citeprefix\cite{MTW}, box~31.2),
so far as we know the $3+1$ field variables for an Eddington-Finkelstein
slice of Schwarzschild spacetime haven't been published before in
the open literature.  For reference we list the key ones here; we
have given a more complete listing elsewhere
(\citeprefix\cite{Thornburg-PhD}, appendix~2).  For a Schwarzschild
spacetime of mass $m$,
\begin{eqnarray}
\alpha
	& = &	\frac{1}{\sqrt{1 + \frac{2m}{r}}}
									\\
\beta^r
	& = &	\frac{2m}{r}
		\frac{1}{1 + \frac{2m}{r}}
									\\
g_{rr}
	& = &	1 + \frac{2m}{r}
									\\
g_{\theta\theta}
	& = &	r^2
									\\
K_{rr}
	& = &	- \frac{2m}{r^2}
		  \frac{1 + \frac{m}{r}}{\sqrt{1 + \frac{2m}{r}}}
							\label{eqn-Schw/EF-Krr}
									\\
K_{\theta\theta}
	& = &	2m
		\frac{1}{\sqrt{1 + \frac{2m}{r}}}
\end{eqnarray}
so that
\begin{eqnarray}
K
	& = &	\frac{2m}{r^2}
		\frac{1 + \frac{3m}{r}}{\left( 1 + \frac{2m}{r} \right)^{3/2}}
									\\
R_{rr}
	& = &	- \frac{2m}{r^3}
		\frac{1}{1 + \frac{2m}{r}}
									\\
R_{\theta\theta}
	& = &	\frac{m}{r}
		\frac{1 + \frac{4m}{r}}{\left( 1 + \frac{2m}{r} \right)^2}
									\\
R
	& = &	\frac{8 m^2}{r^4}
		\frac{1}{\left( 1 + \frac{2m}{r} \right)^2}
									\\
R_{ij} R^{ij}
	& = &	\frac{6 m^2}{r^6}
		\frac{1 + \frac{8}{3} \frac{m}{r}
			+ \frac{16}{3} \frac{m^2}{r^2}}
		     {\left( 1 + \frac{2m}{r} \right)^4}
							\label{eqn-Schw/EF-RR}
\end{eqnarray}

Independent of the slicing, Schwarzschild spacetime has
$\rho = \four\! R = 0$ (vacuum) and
$R_{abcd} R^{abcd} = 48 m^2 / r^6$ (\citeprefix\cite{MTW}, section~31.7).
%
%
\section{Equations for the Spherically Symmetric Scalar Field System}
\label{app-sssf-eqns}

In this appendix we tabulate the equations for all the $3+1$ variables
involved in our initial data algorithm, in terms of the state variables
$A$, $B$, $X$, $Y$, $P$, and $Q$.
%
%
\subsection{Geometric Variables}
\label{app-sssf-eqns/geometric-vars}

The only nonzero (3-)Christoffel symbols are
\begin{mathletters}
\begin{eqnarray}
\Gamma^r_{rr}
	& = &	\thalf \frac{\partial_r A}{A}				\\
\Gamma^r_{\theta\theta}
	& = &	- \thalf \frac{\partial_r B}{A}				\\
\Gamma^r_{\phi\phi}
	& = &	- \thalf \frac{\partial_r B}{A} \sin^2 \theta		\\
\Gamma^\theta_{r\theta}
	=
\Gamma^\theta_{\theta r}
	& = &	\thalf \frac{\partial_r B}{B}				\\
\Gamma^\theta_{\phi\phi}
	& = &	- \sin \theta \, \cos \theta				\\
\Gamma^\phi_{r\phi}
	=
\Gamma^\phi_{\phi r}
	& = &	\thalf \frac{\partial_r B}{B}				\\
\Gamma^\phi_{\theta\phi}
	=
\Gamma^\phi_{\phi\theta}
	& = &	\frac{\cos \theta}{\sin \theta}				
\end{eqnarray}
\end{mathletters}

The 3-Ricci tensor, scalar, and quadratic curvature invariant are
\begin{eqnarray}
R\ij
	& = &	\diag
		\left[
		\begin{array}{c@{\quad}c@{\quad}c}
		R_{rr}
			& R_{\theta\theta}
				& R_{\theta\theta} \sin^2 \theta
		\end{array}
		\right]
									\\
R_{rr}
	& = &	{}
		- \frac{\partial_{rr} B}{B}
		+ \thalf \frac{(\partial_r B)^2}{B^2}
		+ \thalf \frac{(\partial_r A)(\partial_r B)}{A B}
									\\
R_{\theta\theta}
	& = &	{}
		- \thalf \frac{\partial_{rr} B}{A}
		+ 1
		+ \tquarter \frac{(\partial_r A)(\partial_r B)}{A^2}
									\\
R
	& = &	{}
		- 2 \frac{\partial_{rr} B}{A B}
		+ \thalf \frac{(\partial_r B)^2}{A B^2}
		+ \frac{(\partial_r A)(\partial_r B)}{A^2 B}
		+ \frac{2}{B}
									\\
R_{ij} R^{ij}
	& = &	{}
		  \tfrac{3}{2} \frac{(\partial_{rr} B)^2}{A^2 B^2}
		- 2 \frac{\partial_{rr} B}{A B^2}
		- \tfrac{3}{2}
		  \frac{(\partial_r A)(\partial_r B)(\partial_{rr} B)}{A^3 B^2}
		- \frac{(\partial_r B)^2 (\partial_{rr} B)}{A^2 B^3}
							\nonumber	\\
	&   &	{}
		+ \frac{2}{B^2}
		+ \frac{(\partial_r A)(\partial_r B)}{A^2 B^2}
		+ \tfrac{3}{8}
		  \frac{(\partial_r A)^2 (\partial_r B)^2 }{A^4 B^2}
		+ \thalf \frac{(\partial_r A) (\partial_r B)^3 }{A^3 B^3}
		+ \tquarter \frac{(\partial_r B)^4 }{A^2 B^4}
\end{eqnarray}

In spherical symmetry it's relatively easy to locate apparent horizons,
by simply zero-finding on the apparent horizon equation
(\citeprefix\cite{York-1989-in-Frontiers})
\begin{equation}
H \equiv
	\frac{\partial_r B}{\sqrt{A} \, B}
	- 2 \frac{Y}{B}
		= 0
                                                        \label{eqn-horizon}
\end{equation}

[In detail, to locate apparent horizons we first search the grid
function $H$ for sign changes.  For each sign change (distinct
apparent horizon), we construct a local (Lagrange) interpolating
polynomial for the $H$ values in the manner of
appendix~\ref{app-non-smoothness-of-interp-errors}, then use the
\code{ZEROIN} code of \citeprefix\cite{Forsythe-Malcolm-Moler}, chapter~7,
to locate the interpolating polynomial's zero.]

Note that in spherical symmetry any apparent horizon must have
$r_\areal = 2m$.  [To see this,
\footnote{
	 This argument was suggested to the author
	 by N.~\OMurchadha.
	 }
{} observe that in spherical symmetry, the Hawking mass function
agrees with the Misner-Sharp one, and is given by
$m_H = \thalf r_\areal (1 - \tfrac{1}{4} \omega_+ \omega_-)$,
where $\omega_+$ and $\omega_-$ are the null expansions, normalized
to $+2$ in flat spacetime.  An apparent horizon is defined by one or
both of the expansions vanishing, so $m \equiv m_\H = \thalf r_\areal$
there.]
%
%
\subsection{Scalar Field Variables}
\label{app-sssf-eqns/scalar-field-vars}

The $3+1$ energy and momentum densities are
\begin{eqnarray}
4 \pi \rho
	& = &	\thalf \frac{P^2}{A} + \thalf Q^2
									\\
4 \pi j_i
	& = &	\left[
		\begin{array}{c@{\quad}c@{\quad}c}
		- P Q		& 0	& 0	
		\end{array}
		\right]
									\\
4 \pi j^i
	& = &	\left[
		\begin{array}{c@{\quad}c@{\quad}c}
		- \dfrac{P Q}{A} & 0	& 0	
		\end{array}
		\right]
\end{eqnarray}
and the spatial stress-energy tensor and its trace are
\begin{eqnarray}
4 \pi T\ij
	& = &	\diag
		\left[
		\begin{array}{c@{\quad}c@{\quad}c}
		4 \pi T_{rr}
			& 4 \pi T_{\theta\theta}
				& 4 \pi T_{\theta\theta} \sin^2 \theta
		\end{array}
		\right]
									\\
4 \pi T_{rr}
	& = &	\thalf P^2 + \thalf A Q^2
									\\
4 \pi T_{\theta\theta}
	& = &	- \thalf \frac{B}{A} P^2 + \thalf B Q^2
									\\
4 \pi T
	& = &	- \thalf \frac{P^2}{A} + \tfrac{3}{2} Q^2
\end{eqnarray}
%
%
\subsection{Constraints, Mass Functions, and 4-Invariants}
\label{app-sssf-eqns/constraints+mass-fns+4-invariants}

The energy and momentum constraints are
\begin{equation}
C\equiv
	\left(
	R
	+ 2 \frac{Y^2}{B^2}
	+ 4 \frac{X Y}{A B}
	\right)
	-
	\left(
	2 \frac{P^2}{A}
	+ 2 Q^2
	\right)
		= 0
\end{equation}
and
\begin{eqnarray}
C^i
	& = &	\left[
		\begin{array}{c@{\quad}c@{\quad}c}
		C^r	& 0	& 0	
		\end{array}
		\right]
						\label{eqn-sssf/C-i-vector}
									\\
C^r
	& \equiv &
		\left(
		{} \! \!
		- 2 \frac{\partial_r Y}{A B}
		+ \frac{(\partial_r B) Y}{A B^2}
		+ \frac{(\partial_r B) X}{A^2 B}
		\right)
		-
		\left(
		{} \! \!
		- 2 \frac{P Q}{A}
		\right)
			= 0
\end{eqnarray}

In spherical symmetry the mass contained within a given radius is
given by the Misner-Sharp mass function
(\citeprefix\cite{Misner-Sharp-1964-Lagrangian-spherical-collapse};
\citeprefix\cite{MTW}, section~23.5).  As discussed in
appendix~\ref{app-mass-function}, for our purposes this is most
usefully expressed in the form
\begin{equation}
m_\MS =	\thalf \sqrt{B}
	\left(
	1
	- \tfrac{1}{4} \frac{(\partial_r B)^2}{A B}
	+ \frac{Y^2}{B}
	\right)
						\label{eqn-mass-function-MS}
\end{equation}

Notice that since $B \equiv g_{\theta\theta} = O(r^2)$, the leading
term in~\eqref{eqn-mass-function-MS} is $O(r)$ for large $r$.  For our
slices $m_\MS$ is always $O(1)$ outside the scalar field shells, so the
remaining terms in~\eqref{eqn-mass-function-MS} (must) nearly cancel
the leading term.  This cancellation, and thus our computation of
$m_\MS$, is thus quite sensitive to small finite differencing and
roundoff errors at large $r$.

As discussed in section~\ref{sect-diagnostics}, we thus use the
integrated-scalar-field mass function $m_\mu$ defined
by~\eqref{eqn-mass-function-mu}.
\citesprefix\cite{Guven-OMurchadha-1995-constraints-in-spherical-symmetry-I}
\citenumericvsname{defines}{define} $\mu$ as
\begin{equation}
\mu =	\left( \frac{\partial r_\areal}{\partial \ell} \right) \rho
	- \left( r_\areal K_\theta{}^\theta \right) j^r
\end{equation}
where $\ell$ is the proper radial distance in the slice,
\ie{} $d\ell = \sqrt{g_{rr}} \, dr$.  We thus have
\begin{equation}
\mu =	\thalf \frac{\partial_r B}{\sqrt{A} \sqrt{B}} \rho
	- \frac{\sqrt{A}}{\sqrt{B}} Y j^r
								\label{eqn-mu}
\end{equation}

To actually calculate $m_\mu$, we reformulate the
integral~\eqref{eqn-mass-function-mu} in terms of our nonuniform
gridding coordinate~$\wr$ (\cf{}~section~\ref{sect-diagnostics}),
\begin{equation}
m_\mu = m_\MS(r_\min)
	+ \int _{\wr_\min} ^\wr
	       \frac{4 \pi B \mu}{\partial \wr / \partial r}
	       \sqrt{A}
	       \, d\wr
					\label{eqn-mass-function-mu(wr)}
\end{equation}
then use the alternate form of Simpson's rule given by
\citeprefix\cite{Numerical-Recipes-1st-edition},~\eqn{4.1.14}, to do
the numerical integration.  (The standard form of Simpson's rule is
inconvenient here because it only yields results at every 2nd~grid
point.)

As noted in section~\ref{sect-diagnostics}, the 4-Ricci scalar is
given by $\four\! R = 8 \pi (\rho - T)$, so we have
\begin{equation}
\four\! R
	=	2 \frac{P^2}{A} - 2 Q^2
\end{equation}
We compute the 4-Riemann curvature invariant $R_{abcd} R^{abcd}$
via the general expression~\eqref{eqn-four-Riem-Riem}, using the
subexpressions
\begin{eqnarray}
K^i{}_k K^{kj}
	& = &	\diag
		\left[
		\begin{array}{c@{\quad}c@{\quad}c}
		\dfrac{X^2}{A^2}
			& \dfrac{Y^2}{B^2}
				& \dfrac{Y^2}{B^2} \dfrac{1}{\sin^2 \theta}
		\end{array}
		\right]
				\label{eqn-sssf/four-Riem-Riem/first-subexpr}
									\\
K_{ij} K^{jk} K_{kl} K^{li}
	& = &	\frac{X^4}{A^4} + 2 \frac{Y^4}{B^4}
									\\
K_{ij} K^i{}_k K^{kj}
	& = &	\frac{X^3}{A^3} + 2 \frac{Y^3}{B^3}
									\\
K_{ij} K^{ij}
	& = &	\frac{X^2}{A^2} + 2 \frac{Y^2}{B^2}
\end{eqnarray}
The only nonzero components of $\del_i K_{jk}$ and $\del_{[i} K_{j]k}$
are
\begin{mathletters}
\begin{eqnarray}
\del_r K_{rr}
	& = &	\partial_r X - X \frac{\partial_r A}{A}			\\
\del_r K_{\theta\theta}
	& = &	\partial_r Y - Y \frac{\partial_r B}{B}			\\
\del_r K_{\phi\phi}
	& = &	\left[
		\partial_r Y - Y \frac{\partial_r B}{B}
		\right] \sin^2 \theta					\\
\del_\theta K_{r\theta}
	=
\del_\theta K_{\theta r}
	& = &	\thalf
		\left( \frac{X}{A} - \frac{Y}{B} \right)
		\partial_r B						\\
\del_\phi K_{r\phi}
	=
\del_\phi K_{\phi r}
	& = &	\left[
		\thalf
		\left( \frac{X}{A} - \frac{Y}{B} \right)
		\partial_r B
		\right] \sin^2 \theta
\end{eqnarray}
\end{mathletters}
and
\begin{mathletters}
\begin{eqnarray}
\del_{[r} K_{\theta]\theta}
	=
- \del_{[\theta} K_{r]\theta}
	& = &	\partial_r Y - \thalf
			       \left( \frac{X}{A} + \frac{Y}{B} \right)
			       \partial_r B
									\\
\del_{[r} K_{\phi]\phi}
	=
- \del_{[\phi} K_{r]\phi}
	& = &	\left[
		\partial_r Y - \thalf
			       \left( \frac{X}{A} + \frac{Y}{B} \right)
			       \partial_r B
		\right] \sin^2 \theta
\end{eqnarray}
\end{mathletters}
so that
\begin{equation}
\del_{[i} K_{j]k} \, \del^i K^{jk}
	=	\frac{2}{A B^2}
		\left[
		\partial_r Y - \thalf
			       \left( \frac{X}{A} + \frac{Y}{B} \right)
			       \partial_r B
		\right]^2
				\label{eqn-sssf/four-Riem-Riem/last-subexpr}
\end{equation}
%
%
\subsection{York Decomposition Variables and Boundary Conditions}
\label{app-sssf-eqns/York-vars+BCs}

The terms and coefficients in the York equation~\eqref{eqn-York-G}
and its Jacobian~\eqref{eqn-York-G-Jacobian} are
\begin{eqnarray}
\del_i \del^i \Psi
	& = &	\biggl(
		\frac{1}{A}
		\biggr) \partial_{rr} \Psi
		+ \biggl(
		    \frac{\partial_r B}{A B}
		  - \thalf \frac{\partial_r A}{A^2}
		  \biggr) \partial_r \Psi
									\\
F\ij	& = &	\diag
		\left[
		\begin{array}{c@{\quad}c@{\quad}c}
		F_{rr}
			& - \thalf \dfrac{B}{A} F_{rr}
				& - \thalf \dfrac{B}{A} F_{rr} \sin^2 \theta
		\end{array}
		\right]
									\\
F_{rr}	& = &	\biggl(
		\tfrac{2}{3} X - \tfrac{2}{3} \frac{A}{B} Y
		\biggr)
		+ \biggl(
		    \tfrac{2}{3} \partial_r A
		  - \tfrac{2}{3} \frac{A}{B} \partial_r B
		  \biggr) \partial_r \Omega
		+ \biggl(
			  \vphantom{\frac{A}{B}}
		  \tfrac{4}{3} A
		  \biggr) \Omega
									\\
F\ij F\upij
	& = &	\tfrac{3}{2} \biggl( \frac{F_{rr}}{A} \biggr)^2
									\\
{(\Delta_\ell \Omega)}^i
	& = &	\diag
		\left[
		\begin{array}{c@{\quad}c@{\quad}c}
		{(\Delta_\ell \Omega)}^r
			& 0
				& 0
		\end{array}
		\right]
									\\
{(\Delta_\ell \Omega)}^r
	& = &	\biggl(
		\tfrac{4}{3} \frac{1}{A}
		\biggr) \partial_{rr} \Omega
		+ \biggl(
		    \tfrac{2}{3} \frac{\partial_r A}{A^2}
		  + \tfrac{4}{3} \frac{\partial_r B}{A B}
		  \biggr) \partial_r \Omega
								\nonumber
									\\
	&   &	{}
		+ \biggl(
		    \tfrac{2}{3} \frac{\partial_{rr} A}{A^2}
		  - \tfrac{2}{3} \frac{\partial_{rr} B}{A B}
		  - \tfrac{2}{3} \frac{(\partial_r A)^2}{A^3}
		  - \tthird \frac{(\partial_r B)^2}{A B^2}
		  + \frac{(\partial_r A) (\partial_r B)}{A^2 B}
		  \biggr) \Omega
									\\
{(\ell \Omega)}\ij
	& = &	\diag
		\left[
		\begin{array}{c@{\quad}c@{\quad}c}
		{(\ell \Omega)}_{rr}
			& - \thalf \dfrac{B}{A} {(\ell \Omega)}_{rr}
				& - \thalf \dfrac{B}{A} {(\ell \Omega)}_{rr}
				    \sin^2 \theta
		\end{array}
		\right]
									\\
{(\ell \Omega)}_{rr}
	& = &	\biggl(
			\vphantom{\frac{A}{B}}
		\tfrac{4}{3} A
		\biggr) \partial_r \Omega
		+ \biggl(
		    \tfrac{2}{3} \partial_r A
		  - \tfrac{2}{3} \frac{A}{B} \partial_r B
		  \biggr) \Omega
									\\
F\ij {(\ell \Omega)}\upij
	& = &	\tfrac{3}{2} \frac{1}{A^2} F_{rr} {(\ell \Omega)}_{rr}
									\\
\del^i K
	& = &	\diag
		\left[
		\begin{array}{c@{\quad}c@{\quad}c}
		\del^r K
			& 0
				& 0
		\end{array}
		\right]
									\\
\del^r K
	& = &	  \frac{\partial_r X}{A^2}
		+ 2 \frac{\partial_r Y}{A B}
		- \frac{X \partial_r A}{A^3}
		- 2 \frac{Y \partial_r B}{A B^2}
									\\
\del_j E\upij
	& = &	\diag
		\left[
		\begin{array}{c@{\quad}c@{\quad}c}
		\del_j E^{rj}
			& 0
				& 0
		\end{array}
		\right]
									\\
\del_j E^{rj}
	& = &	  \tfrac{2}{3} \frac{\partial_r X}{A^2}
		- \tfrac{2}{3} \frac{\partial_r Y}{A B}
		- \tfrac{2}{3} \frac{X \partial_r A}{A^3}
		- \tfrac{1}{3} \frac{Y \partial_r B}{A B^2}
		+ \frac{X \partial_r B}{A^2 B}
\end{eqnarray}

In spherical symmetry, the outer boundary
conditions~\eqref{eqn-Psi-outer-BC-scalar-Robin}
and~\eqref{eqn-Omega-outer-BC-scalar-Robin} become
\begin{equation}
\partial_r \Psi + \frac{\Psi - 1}{r}
	= O \left( \frac{1}{r^3} \right)
	\approx 0
				\qquad
				\text{(at $r = r_\max$)}
\end{equation}
and
\begin{equation}
\partial_r \Omega + \frac{\Omega}{r}
	= O \left( \frac{1}{r^3} \right)
	\approx 0
				\qquad
				\text{(at $r = r_\max$)}
\end{equation}
respectively.
%
%
\section{The Misner-Sharp Mass Function in Spherical Symmetry}
\label{app-mass-function}

One of our major diagnostics is the Misner-Sharp mass function, which
we use in the form~\eqref{eqn-mass-function-MS}.  Here we outline the
derivation of this result.

\citenameprefix{Misner and Sharp}
\citeprefix\cite{Misner-Sharp-1964-Lagrangian-spherical-collapse}
\citenamesuffix{}
gave the original derivation of the mass function for a spherically
symmetric dynamic spacetime.  However, their results [their
equations~\eqn{6.7} and~\eqn{6.9}] aren't directly usable in a
$3+1$ code such as ours, for two reasons:  First, their results
are expressed in terms of Lagrangian-coordinate (comoving) matter
variables, whereas we use Eulerian (non-comoving) coordinates.
Second, their results are in the form of integrals from the origin
out to a given radius (analogously to~\eqref{eqn-mass-function-mu}
but with lower limits of $r = 0$), whereas with the black hole exclusion
technique the field variables aren't known (and in fact are generally
singular) in the neighbourhood of the origin.

We thus begin instead with the (equivalent) alternate formulation
of the Misner-Sharp mass function given by
\citeprefix\cite{MTW}, section~23.5,
\begin{equation}
m_\MS	=	\half \rbar
		\left(
		1 - \frac{1}{\gbar_{\rbar\rbar}}
		\right)
				\label{eqn-mass-function-MS-Schw-coords}
\end{equation}
where the bars denote the use of Schwarzschild coordinates
$\xbar^a = (\tbar,\rbar,\theta,\phi)$, defined by requiring the
4-metric to be diagonal ($\gbar_{\tbar\rbar} = 0$) and the radial
coordinate to be areal ($\gbar_{\theta\theta} = \rbar^2$).

In our $3+1$ code we permit the coordinates to be arbitrarily
chosen (within the restriction of spherical symmetry), \ie{} we
permit the 4-metric to be non-diagonal and/or the radial coordinate
to be non-areal.  To transform~\eqref{eqn-mass-function-MS-Schw-coords}
back to such generic coordinates, we first define an intermediate
coordinate system $\xhat^a = (t,\rbar,\theta,\phi)$, which uses our
generic time coordinate $t$ but retains the areal radial coordinate
$\rbar$.  From the requirement that $\gbar_{\tbar\rbar} = 0$, a
straightforward calculation shows that
$\gbar_{\rbar\rbar}
	= \ghat_{\rbar\rbar} - (\ghat_{t\rbar})^2 / \ghat_{tt}$,
and hence
\begin{equation}
m_\MS	=	\half \rbar
		\left[
		1 - \frac{1}{\ghat_{\rbar\rbar}}
		  + \left( \frac{\betahat}{\alphahat} \right)^2
		\right]
					\label{eqn-mass-function-MS-alpha-beta}
\end{equation}
[This result appears somewhat problematic, as it has $m$ apparently
depending on the freely specifyable $\xhat^a$-coordinate lapse and
shift $\alphahat$ and $\betahat$.  However, the areal radial
coordinate condition $\ghat_{\theta\theta} = \rbar^2$ actually
fixes the ratio $\betahat / \alphahat$ which enters into $m$
(\ie{} this ratio actually {\em isn't\/} freely specifyable),
so there's no difficulty.]

From the requirement that $\ghat_{\theta\theta} = \rbar^2$
it then follows that
$\partial r / \partial \rbar = 2 \sqrt{B} / \partial_r B$, and hence
\begin{equation}
m_\MS
	=	\thalf \sqrt{B}
		\left(
		1
		- \tfrac{1}{4} \frac{(\partial_r B)^2}{A B}
		+ \frac{Y^2}{B}
		\right)
\end{equation}
which is the desired generic-coordinates form for the mass function.
%
%
\section{Numerical Coordinate Transformations}
\label{app-numerical-coord-xforms}

In $3+1$ numerical relativity one often uses coordinate conditions
which fix the form of the spatial metric.  For example, one may
wish to force the radial coordinate to be \xdefn{areal}
($g_{\theta\theta} = r^2$).  Even if our \Ansatz{} for the base
field variables satisfies such a condition, in general the York
projection operator~\eqref{eqn-York-xform} doesn't preserve it.
As the final step in our overall initial data algorithm
(step~\ref{algorithm-step-coord-xform} in our presentation of
section~\ref{sect-initial-data-algorithm+ansatze}), we thus
optionally apply an explicit numerical coordinate transformation
to the output of the York projection, so as to (\eg{}) convert it
back to an areal radial coordinate.

We actually use a numerical-coordinate-transformation algorithm
somewhat more general than is needed just for this purpose:  our
algorithm actually applies a generic (nonsingular and invertible)
numerically specified radial-coordinate transformation
$r \to \tilde{r} = \tilde{r}(r)$.  (To specify a transformation
to an areal radial coordinate, we simply take
$\tilde{r} = \sqrt{g_{\theta\theta}}$.)

Numerical coordinate transformations are conceptually straightforward,
but somewhat complicated in detail due to the field variables only
being known on the finite difference grid.  In our case the use of
the nonuniform gridding coordinate $\wr$ further complicates matters.
Figure~\ref{fig-cxform-r-flowchart} summarizes our numerical coordinate
transformation algorithm; given this flowchart the operations involved
should be self-evident.  Notice that as well as tensor-transforming
the field variables, the algorithm must also interpolate them from
the $r$ to the $\tilde{r}$ grid.  We use moving local (Lagrange)
polynomial interpolation for this, as discussed in detail in
appendix~\ref{app-non-smoothness-of-interp-errors}.
%
%
\section{Convergence Tests}
\label{app-convergence-tests}

As discussed in section~\ref{sect-sample-results/accuracy}, several
different error sources are present in a $3+1$ code such as ours,
but the dominant one is usually finite differencing truncation error.
As has been forcefully emphasized by
\citenameprefix{Choptuik}
\citesprefix\cite{Choptuik-PhD,Choptuik-1991-consistency,
Choptuik-Goldwirth-Piran-1992-sssf-cmp-3+1-vs-2+2}
\citenamesuffix{},
in such a situation a careful comparison of the errors in
approximating the same physical system at different grid resolutions
can yield a great deal of information about, and very stringent tests
of, the code's numerical performance and correctness.

In the present context, consider some diagnostic $Z$ whose true
(continuum) value $Z^\ast$ is known.  [For example, the true values
of the energy and momentum constraints $C$ and $C^i$ are zero,
and in section~\ref{sect-sample-results/other-perturbations} the
true value of $I / I_{\Schw(m_\total)}$ in the (vacuum) daw5 slice
is~$1$, where $I \equiv R_{abcd} R^{abcd}$.]  Suppose we have a
pair of numerically computed approximate initial data slices, identical
except for having a 2:1~ratio of grid spacings $\Delta \wr$.  As
discussed in detail by \citeprefix\cite{Choptuik-1991-consistency},
if all the field variables are smooth and the code's numerical errors
are dominated by truncation errors from $n$th~order finite differencing,
the numerically computed diagnostics $Z$ must satisfy the Richardson
expansion
\begin{mathletters}
						\label{eqn-error-expansion}
\begin{eqnarray}
Z[\Delta \wr]
	& = &	Z^\ast
		+ (\Delta \wr)^n f
		+ O \bigl( (\Delta \wr)^{n+2} \bigr)
									\\
Z[\Delta \wr/2]
	& = &	Z^\ast
		+ (\Delta \wr/2)^n f
		+ O \bigl( (\Delta \wr)^{n+2} \bigr)
\end{eqnarray}
\end{mathletters}
at each grid point, where $Z[\Delta \wr]$ denotes the numerically
computed $Z$ using the grid spacing $\Delta \wr$, and $f$ is an
$O(1)$ smooth function depending on various high order derivatives
of $Z^\ast$ and the field variables, but {\em not\/} on the grid
resolution.  [We're assuming centered finite differencing here in
writing the higher order terms as $O \bigl( (\Delta \wr)^{n+2} \bigr)$,
otherwise they would only be $O \bigl( (\Delta \wr)^{n+1} \bigr)$.]
Neglecting the higher order terms, i.e.~in the limit of small
$\Delta \wr$, we can eliminate $f$ to obtain a direct relationship
between the code's errors at the two resolutions,
\begin{equation}
\frac{Z[\Delta \wr/2] - Z^\ast}{Z[\Delta \wr] - Z^\ast}
	= \frac{1}{2^n}
						\label{eqn-error-convergence}
\end{equation}
which must hold at each grid point common to the two grids.

We use two methods for testing how well or poorly any particular
set of (finite-resolution) numerical results satisfies the
convergence criterion~\eqref{eqn-error-convergence}:  One method,
used here in figures~\ref{fig-100+200.pqw5.C+C-r.Schw+perturb+final},
\ref{fig-100+200.pqw5i.C.Schw+perturb+final},
\ref{fig-100+200.pw5+qw3.h+MS-mass.relerr},
and~\ref{fig-100+200.daw5c.h+MS-mass+four-Riem-Riem.relerr}, is to
plot the absolute values of the low-resolution and high-resolution
errors ($\bigl| Z[\Delta \wr] - Z^\ast \bigr|$ and
$\bigl| Z[\Delta \wr/2] - Z^\ast \bigr|$ respectively) on a
common {\em logarithmic\/} scale.  If, and given the arguments of
\citeprefix\cite{Choptuik-1991-consistency}, in practice {\em only\/}
if, the error expansions~\eqref{eqn-error-expansion} are valid with
the higher order error terms negligible, i.e.~if and only if the errors
are indeed dominated by the expected $n$th~order finite difference
truncation errors, then the two curves will be identical except for
the high-resolution curve being offset downwards (towards smaller
errors) by a factor of $2^n$ from the low-resolution curve.

Our other method
(\citesprefix\cite{Thornburg-PhD,Thornburg-1996-horizon-finding})
for testing our results against the convergence
criterion~\eqref{eqn-error-convergence}, used here in
figure~\ref{fig-100+200.pqw5.C.conv}, is to plot a scatterplot of
the high-resolution errors $Z[\Delta \wr/2] - Z^\ast$ against the
low-resolution errors $Z[\Delta \wr] - Z^\ast$ at the grid
points common to the two grids.  If, and again in practice
{\em only\/} if, the error expansions~\eqref{eqn-error-expansion}
are valid with the higher order error terms negligible, i.e.~if
and only if the errors are indeed dominated by the expected
$n$th~order finite difference truncation errors, then all the
points in the scatterplot will fall on a line through the origin
with slope $1/2^n$ (${=}\, \tfrac{1}{16}$ for 4th~order convergence).

Comparing these analyses, each has its advantages and disadvantages:
The logarithmic analysis simultaneously tests the convergence
criterion~\eqref{eqn-error-convergence}, and gives a good qualitative
overview of the overall magnitude and spatial variation of the
errors.  Moreover, the use of a logarithmic plot gives this method
excellent dynamic range, \ie{} it can easily test deviations from
the convergence criterion in parts of the grid where the errors
are small, even if elsewhere in the grid the errors are large.
However, this analysis is relatively insensitive to small deviations
from the expected convergence factor.  In contrast, the scatterplot
analysis has excellent sensitivity to small deviations from the
expected convergence factor, but it doesn't give any direct
indication of where in the grid any problems lie, and it has
limited dynamic range unless multiple plots at different scales
are used (as in figure~\ref{fig-100+200.pqw5.C.conv}).

It's important to note that both analyses are applied
{\em pointwise\/}, \ie{} independently at each grid point common
to the two grids.  This makes these analyses much more sensitive
than a simple comparison of gridwise norms, which would tend to
``wash out'' any convergence problems occurring only in a small
subset of the grid points (\eg{} near a boundary).

[One ``false alarm'' problem of this type to which the scatterplot
analysis is particularly subject, is that of boundary finite
differencing effects:  Near a grid boundary, we use off-centered
molecules for finite differencing which, while still 4th~order
in accuracy, are an $O(1)$ factor less accurate than centered
molecules we use in the grid interior.  This in itself doesn't
cause a problem, but the choice of centered \vs{} off-centered
molecules is based on the distance {\em in grid points\/} from
the grid boundary, and so may differ at a near-boundary grid point
between different-resolution grids.  For example, with a 2:1~ratio
of grid resolutions, a point which is 2~grid points from the
boundary (and thus can use a centered 5~point molecule) in the
high-resolution grid, is only 1~grid point from the boundary
(and thus must use an off-centered molecule) in the low-resolution
grid.  The different $O(1)$ factors in the different molecules'
truncation errors then cause spurious apparent changes in the
convergence ratios in~\eqref{eqn-error-convergence} at such
grid points.  This is the cause of the boundary artifacts in
figures~\ref{fig-100+200.pqw5.C+C-r.Schw+perturb+final},
\ref{fig-100+200.pqw5i.C.Schw+perturb+final},
\ref{fig-100+200.pw5+qw3.h+MS-mass.relerr},
and~\ref{fig-100+200.daw5c.h+MS-mass+four-Riem-Riem.relerr}, and
of the outlier points in figure~\ref{fig-100+200.pqw5.C.conv}.]

Finally, note that the parameter $n$, the order of the convergence,
is (or at least should be) known in advance from the form of the
finite differencing scheme.  Thus the factor of $2^n$ convergence
ratio in~\eqref{eqn-error-convergence}, manifested in the vertical
offset between the two plots in the logarithmic analysis, or in the
slope of the line through the origin in the scatterplot analysis,
isn't fitted to the data points, but is rather an a~priori prediction
with {\em no\/} adjustable parameters.  A convergence test of either
type is thus a very strong test of the validity of the overall finite
differencing scheme and the error expansion \eqref{eqn-error-expansion}.
%
%
\section{The Non-Smoothness of Interpolation Errors}
\label{app-non-smoothness-of-interp-errors}

As discussed in appendix~\ref{app-numerical-coord-xforms},
interpolation plays an important role in numerical coordinate
transformation, and it's also important in horizon finding
(\citesprefix\cite{BCSST-1996-horizon-finding,
Thornburg-1996-horizon-finding,ACLMSS-1996-horizon-finding,Huq-PhD}).
In this appendix we discuss some surprising properties of the errors
in interpolation.  For pedagogical convenience we focus only on
(Lagrange) polynomial interpolation, but our results actually hold
for a wide range of interpolation schemes.

To understand the behavior of interpolation errors, it's useful to
consider an artificial example where the data being interpolated
is in fact already known analytically:  Suppose we have a smooth
continuum function $f$ and a set of grid points $\{x_k\}$ in its
domain.  To simplify our discussion we take the grid points $\{x_k\}$
to be uniformly spaced with spacing $\Delta x$, but again this
restriction isn't essential.

Now define an interpolating function $I$ using the function values
$\{f(x_k)\}$ in the usual manner via an $n+1$~point moving local
interpolant:  On each interval $[x_i, x_{i+1}]$ between adjacent
grid points, we set $I$ to the (unique) $n$th-degree interpolating
polynomial $P[i]$ which agrees with the $\{f(x_k)\}$ values
at the $n+1$~nearest-neighbour grid points located symmetrically
about the interval $[x_i, x_{i+1}]$.

Consider the interpolation error $e \equiv I - f$ in taking $I$
as an approximation to $f$, and this error's behavior as the
grid is refined ($\Delta x \to 0$).  It's well known that for
sufficiently smooth $f$, $e = O \bigl( (\Delta x)^{n+1} \bigr)$,
so the error can be made small by refining the grid.  But
{\em the error function $e = e(x)$ is not smooth!\/}  In fact,
even if $f$ is analytic, {\em the error function has a discontinuous
(1st) derivative at each grid point.\/}  This holds regardless of
the interpolation order~$n$.

The reason for this behavior is that although in any grid interval
$[x_k, x_{k+1}]$, $I = P[k]$ is a smooth function (an $n$th-degree
polynomial), $I$ is a {\em different} such polynomial in each separate
interval $[x_k, x_{k+1}]$.  At each grid point the two adjacent
interval's interpolating polynomials agree in their values (so $I$
is continuous there), but not in their derivatives (so $dI/dx$ and
hence $de/dx$ has a jump discontinuity there).  In other words, $e$
is a \defn{bump function}: it vanishes at each grid point, but between
each pair of grid points it rises to a nonzero value, here with
amplitude $O \bigl( (\Delta x)^{n+1} \bigr)$.

Now suppose we approximate derivatives of $f$ by derivatives of the
interpolating function $I$.  Because of $e$'s non-smoothness, the
Richardson-type argument outlined by
\citeprefix\cite{Choptuik-1991-consistency} {\em doesn't\/} apply
here.  Instead, by considering the behavior of the interpolation
process for polynomials $f$ of varying degrees, it's straightforward
to show that for the $p$th derivative the error in this approximation
is
\begin{equation}
\frac{d^p e}{dx^p}
	\equiv \frac{d^p I}{dx^p} - \frac{d^p f}{dx^p}
		= O \bigl( (\Delta x)^{n+1-p} \bigr)
						\label{eqn-interp-dx[p]-error}
\end{equation}
Hence provided $p \leq n$, the error in approximating derivatives
this way can still be made small by refining the grid, but due to
the non-smoothness of the error function the effective order of
convergence is lowered by the number of derivatives taken.

This analysis assumes that all differentiation (\eg{} to compute
$d^p I/dx^p$) is done exactly (analytically).  However, if local
numerical differentiation (finite differencing) is used instead,
then it's easy to show that the error
expansion~\eqref{eqn-interp-dx[p]-error} still holds.

As an example of interpolation errors, consider the function $f$
defined by $f(x) = \exp(\sin(\tfrac{\pi}{2} x))$, and take $n = 3$,
so the local interpolation uses cubic polynomials fitted to
4~nearest-neighbour function values.  Figure~\hbox{\ref{fig-interp-errs}(a)}
shows the error $e$ in this interpolation for grids with spacings
in the 2:1~ratio $\Delta x = 0.1$\,:\,$0.05$, plotted for the
range $[0,0.5]$.  (In each case the grids are enough larger than
this range so the interpolations never encounter the grid endpoints.)
The overall bump-function shape of the error is clearly evident,
with $e$ vanishing at the grid points (the interpolation is exact
there), and rising to a maximum between each pair of grid points.
The amplitudes of the ``bumps'' for the two grid spacings are in
approximately a 16:1~ratio, confirming that we have 4th~order
convergence here, $e = O \bigl( (\Delta x)^4 \bigr)$.

Figure~\hbox{\ref{fig-interp-errs}(b)} shows the error $de/dx$ in
taking $dI/dx$ as an approximation to $df/dx$ for this same example.
$de/dx$ has a jump discontinuity at each grid point, corresponding
to the abrupt change in slope going from one ``bump'' to the next
in $e$ (figure~\hbox{\ref{fig-interp-errs}(a)}).  The amplitudes of
$de/dx$ for the two grid spacings are in approximately an 8:1~ratio,
confirming that we only have 3rd~order convergence here,
$e = O \bigl( (\Delta x)^3 \bigr)$.

Our conclusion from this argument is that by virtue
of~\eqref{eqn-interp-dx[p]-error}, if interpolated data is to be used
as input to further finite differencing computations, then either
the interpolation order should be raised sufficiently to compensate
for the non-smoothness, or Hermite, spline, or other explicitly
smoothness-preserving interpolation schemes should be used.

In our present computational scheme we use moving local polynomial
interpolation in the manner described above.  Our overall computational
scheme is intended to be 4th~order accurate, and we use numerical
differentiation up to 2nd~order (\eg{} in the 3-Ricci tensor), so
we use 5th~order (Lagrange) polynomials for local interpolation
in our numerical coordinate transformation and horizon finding
algorithms.  These interpolants give 6th~order local truncation
errors, so 2nd~numerical derivatives of the interpolation results
should still be 4th~order accurate.
%
%
%
%

\begin{thebibliography}{}

\bibitem[Abrahams and Evans(1992)]{Abrahams-Evans-1992-BH-geon-formation}
Abrahams, A.~M. and Evans, C.~R. (1992).
\newblock Trapping a geon: Black hole formation by an imploding gravitational
  wave.
\newblock {\em Physical Review D}, {\bf 46}(10), R4117--R4121.

\bibitem[Anninos {\em et~al.}(1995)]{ADMSS-1995-BHE}
Anninos, P., Daues, G., \Masso{}, J., Seidel, E., and Suen, W.-M. (1995).
\newblock Horizon boundary condition for black hole spacetimes.
\newblock {\em Physical Review D}, {\bf 51}(10), 5562--5578.

\bibitem[Anninos {\em et~al.}(1996)]{ACLMSS-1996-horizon-finding}
Anninos, P., Camarda, K., Libson, J., \Masso{}, J., Seidel, E., and Suen, W.-M.
  (1996).
\newblock Finding apparent horizons in dynamic 3d numerical spacetimes.

\bibitem[Arnowitt {\em et~al.}(1962)]{ADM-1962}
Arnowitt, R., Deser, S., and Misner, C.~W. (1962).
\newblock The dynamics of general relativity.
\newblock In L.~Witten, editor, {\em Gravitation: An Introduction to Current
  Research}, pages 227--265. Wiley, New York.

\bibitem[Baumgarte {\em et~al.}(1996)]{BCSST-1996-horizon-finding}
Baumgarte, T.~W., Cook, G.~B., Scheel, M.~A., Shapiro, S.~L., and Teukolsky,
  S.~A. (1996).
\newblock Implementing an apparent-horizon finder in three dimensions.
\newblock {\em Physical Review D}, {\bf 54}(8), 4849--4857.

\bibitem[Bernstein and Bartnik(1995)]{Bernstein-Bartnik-1995-sssf}
Bernstein, D.~H. and Bartnik, R. (1995).
\newblock A research in progress report on the evolution of spherically
  symmetric $\sf {SU}(2)$ {E}instein-{Y}ang-{M}ills fields.
\newblock Technical report, University of New England (NSW, Australia).

\bibitem[Bowen(1979a)]{Bowen-1979-momentum-constraint-analytical-soln}
Bowen, J.~M. (1979a).
\newblock General form for the longitudinal momentum of a spherically symmetric
  source.
\newblock {\em General Relativity and Gravitational}, {\bf 11}(3), 227--231.

\bibitem[Bowen(1979b)]{Bowen-PhD}
Bowen, J.~M. (1979b).
\newblock {\em Initial Value Problems on Non-{E}uclidean Topologies}.
\newblock Ph.D. thesis, University of North Carolina at Chapel Hill.

\bibitem[Bowen(1982)]{Bowen-1982-momentum-constraint-analytical-soln}
Bowen, J.~M. (1982).
\newblock General solution for flat-space longitudinal momentum.
\newblock {\em General Relativity and Gravitational}, {\bf 14}(12), 1183--1191.

\bibitem[Bowen(1985)]{Bowen-1985-N-BH-initial-data}
Bowen, J.~M. (1985).
\newblock Inversion symmetric initial data for $n$ charged black holes.
\newblock {\em Annals of Physics}, {\bf 165}, 17--37.

\bibitem[Bowen and York(1980)]{Bowen-York-1980-2BH-initial-data}
Bowen, J.~M. and York, Jr., J.~W. (1980).
\newblock Time-symmetric initial data for black holes and black hole
  collisions.
\newblock {\em Physical Review D}, {\bf 21}(8), 2047--2056.

\bibitem[Bowen {\em et~al.}(1984)]{Bowen-Rauber-York-1984-2BH-initial-data}
Bowen, J.~M., Rauber, J.~D., and York, Jr., J.~W. (1984).
\newblock Two black holes with axisymmetric parallel spins: Initial data.
\newblock {\em Classical and Quantum Gravity}, {\bf 1}, 591--610.

\bibitem[Boyd(1989)]{Boyd}
Boyd, J.~P. (1989).
\newblock {\em {C}hebyshev \& {F}ourier Spectral Methods}, volume~49 of {\em
  Lecture Notes in Engineering}.
\newblock Springer-Verlag, Berlin.

\bibitem[Brill and
  Lindquist(1963)]{Brill-Lindquist-1963-Misner-initial-data-energy}
Brill, D.~R. and Lindquist, R.~W. (1963).
\newblock Interaction energy in geometrostatics.
\newblock {\em Physical Review}, {\bf 131}(1), 471--476.

\bibitem[Centrella {\em et~al.}(1986)]{CSEHT-1986-in-Centrella}
Centrella, J.~M., Shapiro, S.~L., Evans, C.~R., Hawley, J.~F., and Teukolsky,
  S.~A. (1986).
\newblock Test-bed calculations in numerical relativity.
\newblock In J.~M. Centrella, editor, {\em Dynamical Spacetimes and Numerical
  Relativity}, pages 328--344. Cambridge University Press, Cambridge, UK.
\newblock Proceedings of the Workshop on Dynamical Spacetimes and Numerical
  Relativity, Drexel University (Philadelphia, Pennsylvania, USA), 7--11
  October 1985.

\bibitem[Choptuik(1982)]{Choptuik-MSc}
Choptuik, M.~W. (1982).
\newblock {\em A Study of Numerical Techniques for the Initial Value Problem of
  General Relativity}.
\newblock Master's thesis, University of British Columbia.

\bibitem[Choptuik(1986)]{Choptuik-PhD}
Choptuik, M.~W. (1986).
\newblock {\em A Study of Numerical Techniques for Radiative Problems in
  General Relativity}.
\newblock Ph.D. thesis, University of British Columbia.

\bibitem[Choptuik(1991)]{Choptuik-1991-consistency}
Choptuik, M.~W. (1991).
\newblock Consistency of finite-difference solutions of {E}instein's equations.
\newblock {\em Physical Review D}, {\bf 44}(10), 3124--3135.

\bibitem[Choptuik {\em
  et~al.}(1992)]{Choptuik-Goldwirth-Piran-1992-sssf-cmp-3+1-vs-2+2}
Choptuik, M.~W., Goldwirth, D.~S., and Piran, T. (1992).
\newblock A direct comparison of two codes in numerical relativity.
\newblock {\em Classical and Quantum Gravity}, {\bf 9}, 721--750.

\bibitem[Christoudolou(1986a)]{Christoudolou-1986a-sssf}
Christoudolou, D. (1986a).
\newblock {\em Communications in Mathematical Physics}, {\bf 105}, 337.

\bibitem[Christoudolou(1986b)]{Christoudolou-1986b-sssf}
Christoudolou, D. (1986b).
\newblock {\em Communications in Mathematical Physics}, {\bf 106}, 587.

\bibitem[Christoudolou(1987a)]{Christoudolou-1987a-sssf}
Christoudolou, D. (1987a).
\newblock {\em Communications in Mathematical Physics}, {\bf 109}, 591.

\bibitem[Christoudolou(1987b)]{Christoudolou-1987b-sssf}
Christoudolou, D. (1987b).
\newblock {\em Communications in Mathematical Physics}, {\bf 109}, 613--647.

\bibitem[Christoudolou(1991)]{Christoudolou-1991-sssf}
Christoudolou, D. (1991).
\newblock The formation of black holes and singularities in spherically
  symmetric gravitational collapse.
\newblock {\em Communications on Pure and Applied Mathematics}, {\bf 44},
  339--373.

\bibitem[Christoudolou(1993)]{Christoudolou-1993-sssf}
Christoudolou, D. (1993).
\newblock Bounded variation solutions of the spherically symmetric
  {E}instein-scalar field equations.
\newblock {\em Communications on Pure and Applied Mathematics}, {\bf 46},
  1131--1220.

\bibitem[Cook(1990)]{Cook-PhD}
Cook, G.~B. (1990).
\newblock {\em Initial Data for the Two-Body Problem of General Relativity}.
\newblock Ph.D. thesis, University of North Carolina at Chapel Hill.

\bibitem[Cook(1991)]{Cook-1991-2BH-initial-data}
Cook, G.~B. (1991).
\newblock Initial data for axisymmetric black-hole collisions.
\newblock {\em Physical Review D}, {\bf 44}(10), 2983--3000.

\bibitem[Cook {\em et~al.}(1993)]{CCDKMO-1993-3D-2BH-initial-data}
Cook, G.~B., Choptuik, M.~W., Dubal, M.~R., Klasky, S., Matzner, R.~A., and
  Oliveira, S.~R. (1993).
\newblock Three-dimensional initial data for the collision of two black holes.
\newblock {\em Physical Review D}, {\bf 47}(4), 1471--1490.

\bibitem[Dongarra {\em et~al.}(1979)]{LINPACK-book}
Dongarra, J.~J., Moler, C.~B., Bunch, J.~R., and Stewart, G.~W. (1979).
\newblock {\em LINPACK User's Guide}.
\newblock Society for Industrial and Applied Mathematics, Philadelphia.

\bibitem[Dubal(1992)]{Dubal-1992-3D-BH-initial-data}
Dubal, M.~R. (1992).
\newblock Construction of three-dimensional black-hole initial data via
  multiquadrics.
\newblock {\em Physical Review D}, {\bf 45}(4), 1178--1187.

\bibitem[Dubal {\em et~al.}(1992)]{Dubal-Oliveira-Matzner-1992-in-dInverno}
Dubal, M.~R., Oliveira, S.~R., and Matzner, R.~A. (1992).
\newblock Solution of elliptic equations in numerical relativity using
  multiquadrics.
\newblock In R.~d'Inverno, editor, {\em Approaches to Numerical Relativity},
  pages 265--280. Cambridge University Press, Cambridge (UK).
\newblock Proceedings of the International Workshop on Numerical Relativity,
  Southampton University (Southampton, England), 16--20 December 1991.

\bibitem[Einstein and Rosen(1935)]{Einstein-Rosen-1935-bridge}
Einstein, A. and Rosen, N. (1935).
\newblock The particle problem in the general theory of relativity.
\newblock {\em Physical Review}, {\bf 48}(10), 73--77.

\bibitem[Forsythe {\em et~al.}(1977)]{Forsythe-Malcolm-Moler}
Forsythe, G.~E., Malcolm, M.~A., and Moler, C.~B. (1977).
\newblock {\em Computer Methods for Mathematical Computations}.
\newblock Prentice-Hall, Englewood Cliffs.

\bibitem[Goldwirth and Piran(1987)]{Goldwirth-Piran-1987-sssf-2+2}
Goldwirth, D.~S. and Piran, T. (1987).
\newblock {\em Physical Review D}, {\bf 36}, 3575--3581.

\bibitem[Goldwirth {\em et~al.}(1989)]{Goldwirth-Ori-Piran-1989-in-Frontiers}
Goldwirth, D.~S., Ori, A., and Piran, T. (1989).
\newblock Cosmic censorship and numerical relativity.
\newblock In C.~R. Evans, L.~S. Finn, and D.~W. Hobill, editors, {\em Frontiers
  in Numerical Relativity}, pages 414--435. Cambridge University Press,
  Cambridge (UK).
\newblock Proceedings of the International Workshop on Numerical Relativity,
  University of Illinois at Urbana-Champaign (Urbana-Champaign, Illinois, USA),
  9--13 May 1988.

\bibitem[\Gomez{} {\em et~al.}(1996)]{GLPW-1996-sssf-3+1-and-2+2}
\Gomez{}, R., Laguna, P., Papadopoulos, P., and Winicour, J. (1996).
\newblock {C}auchy-characteristic evolution of {E}instein-{K}lein-{G}ordon
  systems.
\newblock {\em Physical Review D}, {\bf 54}(8), 4719--4727.

\bibitem[Guven and
  \OMurchadha(1995)]{Guven-OMurchadha-1995-constraints-in-spherical-symmetry-I}
Guven, J. and \OMurchadha, N. (1995).
\newblock The constraints in spherically symmetric classical general
  relativity: {I}. optical scalars, foliations, bounds on the configuration
  space variables, and the positivity of the quasilocal mass.
\newblock {\em Physical Review D}, {\bf 52}(2), 758--775.

\bibitem[\Hamade{} and Stewart(1996)]{Hamade-Stewart-1996-sssf-2+2}
\Hamade{}, R.~S. and Stewart, J.~M. (1996).
\newblock The spherically symmetric collapse of a massless scalar field.
\newblock {\em Classical and Quantum Gravity}, {\bf 13}, 497--512.

\bibitem[Huq(1996)]{Huq-PhD}
Huq, M.~F. (1996).
\newblock {\em Apparent Horizon Location in Numerical Spacetimes}.
\newblock Ph.D. thesis, University of Texas at Austin.

\bibitem[Isenberg(1979)]{Isenberg-PhD}
Isenberg, J.~A. (1979).
\newblock {\em The Construction of Spacetimes from Initial Data}.
\newblock Ph.D. thesis, University of Maryland.

\bibitem[Kulkarni(1984a)]{Kulkarni-1984-2BH-initial-data}
Kulkarni, A.~D. (1984a).
\newblock Extrinsic curvature for the two-black-hole problem.
\newblock {\em General Relativity and Gravitation}, {\bf 17}(4), 301--310.

\bibitem[Kulkarni(1984b)]{Kulkarni-1984-N-BH-initial-data}
Kulkarni, A.~D. (1984b).
\newblock Time-asymmetric initial data for the $n$ black hole problem in
  general relativity.
\newblock {\em Journal of Mathematical Physics}, {\bf 25}(4), 1028--1034.

\bibitem[Kulkarni {\em
  et~al.}(1983)]{Kulkarni-Shepley-York-1983-N-BH-initial-data}
Kulkarni, A.~D., Shepley, L.~C., and York, Jr., J.~W. (1983).
\newblock Initial data for $n$ black holes.
\newblock {\em Physics Letters}, {\bf 96A}(5), 228--230.

\bibitem[Marsa(1995)]{Marsa-PhD}
Marsa, R.~L. (1995).
\newblock {\em Radiative Problems in Black Hole Spacetimes}.
\newblock Ph.D. thesis, University of Texas at Austin.

\bibitem[Marsa and Choptuik(1996)]{Marsa-Choptuik-1996-sssf}
Marsa, R.~L. and Choptuik, M.~W. (1996).
\newblock Black hole--scalar field interactions in spherical symmetry.
\newblock {\em Physical Review D}, {\bf 54}(8), 4929--4943.

\bibitem[Matzner {\em
  et~al.}(1998)]{Matzner-Huq-Shoemaker-1998-N-BH-initial-data}
Matzner, R.~A., Huq, M.~F., and Shoemaker, D. (1998).
\newblock Initial data and coordinates for multiple black hole systems.

\bibitem[Misner(1960)]{Misner-1960-initial-data}
Misner, C.~W. (1960).
\newblock Wormhole initial conditions.
\newblock {\em Physical Review}, {\bf 118}(4), 1110--1111.

\bibitem[Misner and
  Sharp(1964)]{Misner-Sharp-1964-Lagrangian-spherical-collapse}
Misner, C.~W. and Sharp, D.~H. (1964).
\newblock Relativistic equations for adiabatic, spherically symmetric
  gravitational collapse.
\newblock {\em Physical Review B}, {\bf 136}(2), 571--576.

\bibitem[Misner {\em et~al.}(1973)]{MTW}
Misner, C.~W., Thorne, K.~S., and Wheeler, J.~A. (1973).
\newblock {\em Gravitation}.
\newblock W. H. Freeman, San Francisco.

\bibitem[\OMurchadha{}(1992)]{OMurchadha-1992-in-dInverno}
\OMurchadha{}, N. (1992).
\newblock Boundary conditions for the momentum constraint.
\newblock In R.~d'Inverno, editor, {\em Approaches to Numerical Relativity},
  pages 83--93. Cambridge University Press, Cambridge (UK).
\newblock Proceedings of the International Workshop on Numerical Relativity,
  Southampton University (Southampton, England), 16--20 December 1991.

\bibitem[\OMurchadha{} and
  York(1974a)]{OMurchadha-York-1974-York-decomposition-I}
\OMurchadha{}, N. and York, Jr., J.~W. (1974a).
\newblock Initial-value problem of general relativity: {I} -- general
  formulation and physical interpretation.
\newblock {\em Physical Review D}, {\bf 10}(2), 428--436.

\bibitem[\OMurchadha{} and
  York(1974b)]{OMurchadha-York-1974-York-decomposition-II}
\OMurchadha{}, N. and York, Jr., J.~W. (1974b).
\newblock Initial-value problem of general relativity: {II} -- stability of
  solutions of the initial-value equations.
\newblock {\em Physical Review D}, {\bf 10}(2), 437--446.

\bibitem[Oohara and Nakamura(1989)]{Oohara-Nakamura-1989-in-Frontiers}
Oohara, K. and Nakamura, T. (1989).
\newblock Three dimensional initial data of numerical relativity.
\newblock In C.~R. Evans, L.~S. Finn, and D.~W. Hobill, editors, {\em Frontiers
  in Numerical Relativity}, pages 74--88. Cambridge University Press, Cambridge
  (UK).
\newblock Proceedings of the International Workshop on Numerical Relativity,
  University of Illinois at Urbana-Champaign (Urbana-Champaign, Illinois, USA),
  9--13 May 1988.

\bibitem[Press {\em et~al.}(1986)]{Numerical-Recipes-1st-edition}
Press, W.~H., Flannery, B.~P., Teukolsky, S.~A., and Vetterling, W.~T. (1986).
\newblock {\em Numerical Recipes: The Art of Scientific Computing}.
\newblock Cambridge University Press, Cambridge (UK) and New York, 1st edition.

\bibitem[Rauber(1985)]{Rauber-PhD}
Rauber, J.~D. (1985).
\newblock {\em Initial Data for Black Hole Collisions}.
\newblock Ph.D. thesis, University of North Carolina at Chapel Hill.

\bibitem[Rauber(1986)]{Rauber-1986-in-Centrella}
Rauber, J.~D. (1986).
\newblock Initial data for black hole collisions.
\newblock In J.~M. Centrella, editor, {\em Dynamical Spacetimes and Numerical
  Relativity}, pages 304--327. Cambridge University Press, Cambridge, UK.
\newblock Proceedings of the Workshop on Dynamical Spacetimes and Numerical
  Relativity, Drexel University (Philadelphia, Pennsylvania, USA), 7--11
  October 1985.

\bibitem[Scheel {\em
  et~al.}(1995a)]{Scheel-Shapiro-Teukolsky-1995a-BHE-Brans-Dicke}
Scheel, M.~A., Shapiro, S.~L., and Teukolsky, S.~A. (1995a).
\newblock Collapse to black holes in {B}rans-{D}icke theory: {I}. horizon
  boundary conditions for dynamical spacetimes.
\newblock {\em Physical Review D}, {\bf 51}, 4208--4235.

\bibitem[Scheel {\em
  et~al.}(1995b)]{Scheel-Shapiro-Teukolsky-1995b-BHE-Brans-Dicke}
Scheel, M.~A., Shapiro, S.~L., and Teukolsky, S.~A. (1995b).
\newblock Collapse to black holes in {B}rans-{D}icke theory: {II}. comparison
  with general relativity.
\newblock {\em Physical Review D}, {\bf 51}, 4236.

\bibitem[Scheel {\em et~al.}(1997)]{SBCST-1997-3D-BHE-evolution}
Scheel, M.~A., Baumgarte, T.~W., Cook, G.~B., Shapiro, S.~L., and Teukolsky,
  S.~A. (1997).
\newblock Numerical evolution of black holes with a hyperbolic formulation of
  general relativity.
\newblock {\em Physical Review D}, {\bf 56}(10), 6320--6335.

\bibitem[Seidel and Suen(1992)]{Seidel-Suen-1992-BHE}
Seidel, E. and Suen, W.-M. (1992).
\newblock Towards a singularity-proof scheme in numerical relativity.
\newblock {\em Physical Review Letters}, {\bf 69}(13), 1845--1848.

\bibitem[Shapiro and Teukolsky(1986)]{Shapiro-Teukolsky-1986-in-Centrella}
Shapiro, S.~L. and Teukolsky, S.~A. (1986).
\newblock Relativistic stellar dynamics on the computer.
\newblock In J.~M. Centrella, editor, {\em Dynamical Spacetimes and Numerical
  Relativity}, pages 74--100. Cambridge University Press, Cambridge, UK.
\newblock Proceedings of the Workshop on Dynamical Spacetimes and Numerical
  Relativity, Drexel University (Philadelphia, Pennsylvania, USA), 7--11
  October 1985.

\bibitem[Smarr(1984)]{Smarr-1984-in-GR10}
Smarr, L.~L. (1984).
\newblock Computational relativity: Numerical and algebraic approaches (report
  of workshop {A}4).
\newblock In B.~Bertotti, F.~de~Felice, and A.~Pascolini, editors, {\em General
  Relativity and Gravitation: Invited Papers and Discussion Reports of the 10th
  International Conference on General Relativity and Gravitation, Padua, July
  3--8 1983}, pages 163--183. Reidel, Dordrecht (Netherlands).

\bibitem[Thornburg(1985)]{Thornburg-MSc}
Thornburg, J. (1985).
\newblock {\em Coordinates and Boundary Conditions for the General Relativistic
  Initial Data Problem}.
\newblock Master's thesis, University of British Columbia.

\bibitem[Thornburg(1987)]{Thornburg-1987-2BH-initial-data}
Thornburg, J. (1987).
\newblock Coordinates and boundary conditions for the general relativistic
  initial data problem.
\newblock {\em Classical and Quantum Gravity}, {\bf 4}, 1119--1131.

\bibitem[Thornburg(1991)]{Thornburg-1991-BHE-talk}
Thornburg, J. (1991).
\newblock Numerical relativity in black hole spacetimes.

\bibitem[Thornburg(1993)]{Thornburg-PhD}
Thornburg, J. (1993).
\newblock {\em Numerical Relativity in Black Hole Spacetimes}.
\newblock Ph.D. thesis, University of British Columbia.

\bibitem[Thornburg(1996)]{Thornburg-1996-horizon-finding}
Thornburg, J. (1996).
\newblock Finding apparent horizons in numerical relativity.
\newblock {\em Physical Review D}, {\bf 54}(8), 4899--4918.

\bibitem[Thornburg(1998)]{Thornburg-1998-sssf-evolution}
Thornburg, J. (1998).
\newblock A $3+1$ computational scheme for spherically symmetric dynamic black
  hole spacetimes -- {II}: Time evolution.
\newblock in preparation, to be submitted to Physical Review D.

\bibitem[Wald(1984)]{Wald}
Wald, R.~M. (1984).
\newblock {\em General Relativity}.
\newblock University of Chicago Press, Chicago.

\bibitem[York(1971)]{York-1971-York-decomposition}
York, Jr., J.~W. (1971).
\newblock Gravitational degrees of freedom and the initial-value problem.
\newblock {\em Physical Review Letters}, {\bf 26}(26), 1656--1658.

\bibitem[York(1972)]{York-1972-York-decomposition}
York, Jr., J.~W. (1972).
\newblock Role of conformal three-geometry in the dynamics of gravitation.
\newblock {\em Physical Review Letters}, {\bf 28}(16), 1082--1085.

\bibitem[York(1973)]{York-1973-York-decomposition}
York, Jr., J.~W. (1973).
\newblock Conformally invariant orthogonal decomposition of symmetric tensors
  on {R}iemannian manifolds and the initial value problem of general
  relativity.
\newblock {\em Journal of Mathematical Physics}, {\bf 14}(4), 456--464.

\bibitem[York(1979)]{York-1979-in-Yellow}
York, Jr., J.~W. (1979).
\newblock Kinematics and dynamics of general relativity.
\newblock In L.~L. Smarr, editor, {\em Sources of Gravitational Radiation},
  pages 83--126. Cambridge University Press, Cambridge, UK.

\bibitem[York(1980)]{York-1980-in-Taub-festschrift}
York, Jr., J.~W. (1980).
\newblock Energy and momentum of the gravitational field.
\newblock In F.~J. Tipler, editor, {\em Essays in General Relativity: A
  Festschrift for {A}braham {T}aub}, pages 39--58. Academic Press, New York.

\bibitem[York(1983)]{York-1983-in-Red}
York, Jr., J.~W. (1983).
\newblock The initial value problem and dynamics.
\newblock In N.~Deruelle and T.~Piran, editors, {\em Gravitational Radiation},
  pages 175--201. North-Holland, Amsterdam.

\bibitem[York(1984)]{York-1984-N-BH-initial-data}
York, Jr., J.~W. (1984).
\newblock Initial data for $n$ black holes.
\newblock {\em Physica A}, {\bf 124}, 629--638.

\bibitem[York(1985)]{York-1985-in-Centrella-LeBlank-Bowers}
York, Jr., J.~W. (1985).
\newblock Spacetime engineering.
\newblock In J.~M. Centrella, J.~M. LeBlanc, and R.~L. Bowers, editors, {\em
  Numerical Astrophysics}, pages 176--189. Jones and Bartlett, Boston.

\bibitem[York(1989)]{York-1989-in-Frontiers}
York, Jr., J.~W. (1989).
\newblock Initial data for collisions of black holes and other gravitational
  miscellany.
\newblock In C.~R. Evans, L.~S. Finn, and D.~W. Hobill, editors, {\em Frontiers
  in Numerical Relativity}, pages 89--109. Cambridge University Press,
  Cambridge (UK).
\newblock Proceedings of the International Workshop on Numerical Relativity,
  University of Illinois at Urbana-Champaign (Urbana-Champaign, Illinois, USA),
  9--13 May 1988.

\bibitem[York and Piran(1982)]{York-Piran-1982-in-Schild-lectures}
York, Jr., J.~W. and Piran, T. (1982).
\newblock The initial value problem and beyond.
\newblock In R.~A. Matzner and L.~C. Shepley, editors, {\em Spacetime and
  Geometry: The {A}lfred {S}child Lectures}, pages 147--176. University of
  Texas Press, Austin (Texas).

\end{thebibliography}

%
%
%
\clearpage
%
%
\begin{figure}
%
%
\caption[$\wr$ Nonuniform Gridding Coordinate]
	{
	This figure shows the behavior of our nonuniform gridding
	coordinate $\wr$.  Part~(a) shows the grid spacing
	$\Delta r$ for grids with resolutions of (top to bottom)
	$\Delta \wr = 0.01$, $0.005$, and~$0.0025$.  The diagonal
	dashed lines labeled along the top and right of the
	figure show lines of constant relative grid spacing
	$\Delta r/r$.  Part~(b) shows the actual $\wr$ coordinate.
	In both parts of the figure the vertical dashed lines
	show the outer grid boundaries for $\wr_\max = 4$, $10$,
	and~$30$.
	}
\label{fig-mixed-210-coord}
%
%
\end{figure}
%
%
\begin{figure}
%
%
\caption[Conformal Factor and Vector Potential for 200.pqw5 Slice]
	{
	This figure shows the York-transformation conformal factor
	$\Psi$ and vector potential $\Omega$ for the 200.pqw5 slice.
	The vertical dashed line at $r_\init = 2$ shows the
	input horizon position.
	}
\label{fig-200.pqw5-h+Psi+Omega}
%
%
\end{figure}
%
\begin{figure}
%
%
\caption[Relative 3-Metric and Extrinsic Curvature Components
	 for the 200.pqw5 Slice]
	{
	This figure shows the relative 3-metric and
	extrinsic curvature components $A \equiv g_{rr}$,
	$B \equiv g_{\theta\theta}$, $X \equiv K_{rr}$, and
	$Y \equiv K_{\theta\theta}$, for the 200.pqw5 slice.
	(Note that $B / B_\Schw \equiv 1$ by the definition
	of an areal radial coordinate.)  The vertical dashed
	line just inside $r = 2$ shows the horizon position.
	}
\label{fig-200.pqw5-h+ABXY.ratio}
%
%
\end{figure}
%
\begin{figure}
%
%
\caption[Relative Extrinsic Curvature, 3-Ricci Curvature, and
	 3-Ricci Quadratic Curvature Invariant for 200.pqw5 Slice]
	{
	This figure shows the relative 3-invariants $K$, $R$,
	and $J \equiv R_{ij} R^{ij}$, for the 200.pqw5 slice.
	The horizontal dashed line shows the ``zero axis''
	$K/K_\Schw = R/R_\Schw = J/J_\Schw = 1$.  The vertical
	dashed line just inside $r = 2$ shows the horizon position.
	}
\label{fig-200.pqw5.h+K+R+RR.ratio}
%
%
\end{figure}
%
\begin{figure}
%
%
\caption[Scalar Field and Mass Distributions for 200.pqw5 Slice]
	{
	This figure shows the scalar field and mass distributions
	for the 200.pqw5 slice.  Part~(a) shows the full data range,
	with logarithmic scales for most of the variables, while
	part~(b) shows only the inner part of the grid, with linear
	scales for all variables.  (In part~(b) the left axis uses
	a different scale for $|\four\! R|$ than for $4 \pi B \rho$
	and $|P|$).  In both parts of the figure the vertical dashed
	line just inside $r = 2$ shows the horizon position.
	}
\label{fig-200.pqw5-h+scalar-field+mass}
%
%
\end{figure}
%
%
\begin{figure}
%
%
\caption[Mass-Relative 4-Riemann Curvature Invariant for the 200.pqw5 Slice]
	{
	This figure shows the mass-relative 4-Riemann curvature
	invariant $I \equiv R_{abcd} R^{abcd}$ for the 200.pqw5 slice.
	The vertical dashed line just inside $r = 2$ shows the
	horizon position.  Notice that to within the accuracy of
	the plot, $I / I_{\Schw(m(r))} = 1$ everywhere outside
	the scalar field shell.
	}
\label{fig-200.pqw5-h+four-Riem-Riem.mass-ratio}
%
%
\end{figure}
%
\begin{figure}
%
%
\caption[Mass-Relative Extrinsic Curvature, 3-Ricci Curvature, and
	 3-Ricci Quadratic Curvature Invariant for 200.pqw5 Slice]
	{
	This figure shows the mass-relative 3-invariants $K$,
	$R$, and $J \equiv R_{ij} R^{ij}$, for the 200.pqw5
	slice.  The upper and lower horizontal dashed lines show
	the ``zero axes'' $K/K_{\Schw(m(r))} = 1$ (left scale) and
	$R/R_{\Schw(m(r))} = J/J_{\Schw(m(r))} = 1$ (right scale)
	respectively.  The vertical dashed line just inside
	$r = 2$ shows the horizon position.
	}
\label{fig-200.pqw5.h+K+R+RR.mass-ratio}
%
%
\end{figure}
%
%
\begin{figure}
%
%
\caption[Energy and Momentum Constraints for 100.pqw5 and 200.pqw5 Slices]
	{
	This figure shows the magnitudes of the numerically
	computed energy constraint $C$ (part~(a)), and
	momentum constraint $C^r$ (part~(b)), at various points
	in our initial data algorithm for the 100.pqw5 and
	200.pqw5 slices.  $|C_\Schw|$ ($|C^r_\Schw|$) is the
	magnitude of the energy (momentum) constraint of the
	initial Schwarzschild slice before applying the perturbation
	(step~\ref{algorithm-step-analytical-BH-slice}
	in our overall initial data algorithm of
	section~\ref{sect-initial-data-algorithm+ansatze});
	$|C_\perturb|$ ($|C^r_\perturb|$) is the magnitude of
	the energy (momentum) constraint after applying the
	perturbation (step~\ref{algorithm-step-perturbation});
	and $|C_\final|$ ($|C^r_\final|$) is the magnitude of
	the final energy (momentum) constraint after applying
	the York algorithm (step~\ref{algorithm-step-York-projection})
	and the numerical coordinate transformation back to an areal
	radial coordinate (step~\ref{algorithm-step-coord-xform}).
	The scale bars show a factor of~$16$ in $|C|$ ($|C^r|$),
	for comparison with the vertical spacing between the
	100.pqw5 and 200.pqw5 curves.  For the 200.pqw5
	$|C_\perturb|$ curve, only every 5th~grid point
	is plotted.
	}
\label{fig-100+200.pqw5.C+C-r.Schw+perturb+final}
%
%
\end{figure}
%
\begin{figure}
%
%
\caption[$C_\final$ Convergence Plot for 100.pqw5 and 200.pqw5 Slices]
	{
	This figure shows a scatterplot of $C_\final$
	between the 100.pqw5 and 200.pqw5 slices.  Part~(a)
	shows all the $C_\final$ values, while part~(b)
	shows an expanded view of the central region
	(the grid-interior points) of part~(a).  In each part
	the line through the origin has slope $\tfrac{1}{16}$;
	As discussed in appendix~\ref{app-convergence-tests},
	for 4th~order convergence we expect all the points
	(except for grid-boundary outliers) to fall on the line.
	}
\label{fig-100+200.pqw5.C.conv}
%
%
\end{figure}
%
\begin{figure}
%
%
\caption[Embedding and Curvature Diagnostics
	 for 100o30.pqw5 and 200o30.pqw5 Slices]
	{
	This figure shows various embedding and curvature
	diagnostics for the outermost part of the grid in
	the 100o30.pqw5 slice (part~(a)), and the 200o30.pqw5
	slice (part~(b)).  Each part of the figure shows
	the relative $K$; the mass-relative $J \equiv R_{ij} R^{ij}$
	and $I \equiv R_{abcd} R^{abcd}$; the relative $R$;
	and the relative $R_\fit$ (an empirical least-squares fit
	to $R$) perturbed by simulated roundoff errors in the
	3-metric components, with $\varepsilon = 6{\times}10^{-19}$.
	One outlier point in part~(b), that for $R/R_\Schw$ at
	the outermost grid point, falls outside the plot's range
	and is omitted.  (This point is anomalous due to boundary
	finite differencing effects.)  Note that the plot key
	(the choice of line and point types used for the different
	diagnostics) differs between this and the other figures.
	}
\label{fig-100o30+200o30.pqw5-outer-curvature-ratio+mass-ratio}
%
%
\end{figure}
%
%
\begin{figure}
%
%
\caption[Energy Constraint $C$ for 100.pqw5i and 200.pqw5i Slices]
	{
	This figure shows the magnitude of the numerically
	computed energy constraint, $|C|$, at various points
	in our initial data algorithm for the 100.pqw5i and
	200.pqw5i slices, analogously to
	figure~\ref{fig-100+200.pqw5.C+C-r.Schw+perturb+final}(a)
	for the 100.pqw5 and 200.pqw5 slices.
	$|C_\Schw|$ is the magnitude of
	the energy constraint of the initial Schwarzschild
	slice before applying the perturbation
	(step~\ref{algorithm-step-analytical-BH-slice}
	in our overall initial data algorithm of
	section~\ref{sect-initial-data-algorithm+ansatze});
	$|C_\perturb|$ is the magnitude of the energy
	constraint after applying the perturbation
	(step~\ref{algorithm-step-perturbation}), and
	$|C_\final|$ is the magnitude of the final energy
	constraint after applying the York algorithm
	(step~\ref{algorithm-step-York-projection}) and
	the numerical coordinate transformation back to an areal
	radial coordinate (step~\ref{algorithm-step-coord-xform}).
	The scale bar shows a factor of~$16$ in $|C|$, for
	comparison with the vertical spacing between the
	100.pqw5i and 200.pqw5i curves.  For the 200.pqw5i
	$|C_\perturb|$ curve, only every 5th~grid point is
	plotted.
	}
\label{fig-100+200.pqw5i.C.Schw+perturb+final}
%
%
\end{figure}
%
%
\begin{figure}
%
%
\caption[Scalar Field and Mass Distributions for
	 200.pqw5b and 200.pqw5c Slices]
	{
	This figure shows the scalar field and mass distributions
	for the 200.pqw5b (part~(a)) and 200.pqw5c (part~(b)) slices,
	analogously to figure~\ref{fig-200.pqw5-h+scalar-field+mass}(a)
	for the 200.pqw5 slice.  In both parts of the
	figure the left scale is identical to that of
	figure~\ref{fig-200.pqw5-h+scalar-field+mass}(a);
	the right scale differs for each part.  In part~(a)
	the vertical dashed line just inside $r = 2$ shows
	the horizon position.  (As discussed in the text,
	there is no horizon within the numerical grid in
	the 200.pqw5c slice plotted in part~(b).)
	}
\label{fig-200.pqw5b+pqw5c-h+scalar-field+mass}
%
%
\end{figure}
%
\begin{figure}
%
%
\caption[Scalar Field and Mass Distributions for 400.pqw1 Slice]
	{
	This figure shows the scalar field and mass
	distributions for the 400.pqw1 slice, analogously to
	figure~\ref{fig-200.pqw5-h+scalar-field+mass}(a)
	for the 200.pqw5 slice and
	figure~\ref{fig-200.pqw5b+pqw5c-h+scalar-field+mass}
	for the 200.pqw5b and 200.pqw5c slices.
	Part~(a) shows the full data range, while
	part~(b) shows only the inner part of the grid.
	In both parts of the figure the vertical dashed
	line just inside $r = 2$ shows the horizon position.
	}
\label{fig-400.pqw1-h+scalar-field+mass}
%
%
\end{figure}
%
\begin{figure}
%
%
\caption[Relative Deviation of Misner-Sharp
	 from Integrated--Scalar-Field Mass Function
	 for 100.pw5+qw3 and 200.pw5+qw3 Slices]
	{
	This figure shows the relative deviation $|m_\MS/m_\mu - 1|$ 
	of the Misner-Sharp mass function $m_\MS$ from the
	integrated--scalar-field mass function $m_\mu$, for the
	100.pw5+qw3 and 200.pw5+qw3 slices.  The scale bar shows a
	factor of~$16$ in the deviations, for comparison with the
	vertical spacing between the 100.pw5+qw3 and 200.pw5+qw3
	curves.  The vertical dashed line just inside $r = 2$
	shows the horizon position.
	}
\label{fig-100+200.pw5+qw3.h+MS-mass.relerr}
%
%
\end{figure}
%
\begin{figure}
%
%
\caption[Relative Errors in Misner-Sharp Mass Function
	 and 4-Riemann Curvature Invariant
	 for 100.daw5c and 200.daw5c Slices]
	{
	This figure shows the relative errors
	$|m_\MS/m_\total - 1|$ and $|I/I_{\Schw(m_\total)} - 1|$
	in the Misner-Sharp mass function $m_\MS$ and the 4-Riemann
	curvature invariant $I \equiv R_{abcd} R^{abcd}$, for the
	100.daw5c and 200.daw5c slices, where $m_\total$ is the
	total mass of the slice.  The scale bar shows a factor
	of~$16$ in the errors, for comparison with the vertical
	spacing between the 100.daw5c and 200.daw5c curves.
	The vertical dashed line just outside $r = 2$ shows
	the horizon position.
	}
\label{fig-100+200.daw5c.h+MS-mass+four-Riem-Riem.relerr}
%
%
\end{figure}
%
%
\begin{figure}
%
%
\caption[Numerical Coordinate Transformation Algorithm]
	{
	This figure shows the overall data flow in our
	numerical coordinate transformation algorithm.
	Oval boxes denote grid functions, rectangular boxes
	denote operations performed on grid functions,
	solid arrows show data flows presently used in our algorithm,
	and the dashed arrow shows an additional data flow which
	would be needed to transform contravariant tensors in
	addition to covariant ones.  $f$~@~$\wr$ and
	$f$~@~$\tilde{\wr}$ denote a grid function $f$ defined
	on the $\wr$ and $\tilde{\wr}$ grids respectively.  
	}
\label{fig-cxform-r-flowchart}
%
%
\end{figure}
%
%
\begin{figure}
%
%
\caption[Example of Interpolation Errors]
	{
	This figure shows the interpolation error
	$e \equiv I - f$ (part~(a)) and its derivative
	$de/dx \equiv dI/dx - df/dx$ (part~(b)),
	for the function $f$ defined by
	$f(x) = \exp(\sin(\tfrac{\pi}{2} x))$,
	cubically interpolated to grids with spacings
	$\Delta x = 0.1$ and~$0.05$.  The plot key for
	part~(a) also applies to part~(b).
	}
\label{fig-interp-errs}
%
%
\end{figure}
%
%
%
\clearpage
%
\tighten
%
%
\begin{table}
%
%
%
\caption[Parameters for Test Slices]
	{
	This table gives various parameters for our test initial
	data slices.  For each slice the table shows a descriptive
	name, the spatial grid resolution and size, the initial
	perturbation (step~\ref{algorithm-step-perturbation}
	in our overall initial data algorithm of
	section~\ref{sect-initial-data-algorithm+ansatze}),
	an approximate Gaussian fit to the slice's radial
	scalar field density $4 \pi B \rho$, the horizon's areal
	radius, and a summary of the slice's mass distribution
	(broken down into black hole, scalar field, and total mass).
	}
\label{tab-test-slice-pars}
%
%
\begin{center}
%
%
\mathsurround=0em
%
%
%
\begin{tabular}{llcccccc@{\,}c@{\,}c@{\,}c@{\,}c@{}c@{}c@{}c@{}c@{}c@{\quad}c@{}c@{}c@{}c@{}c@{}c@{\quad}cccccc}
%
\nop{Name}
	& \nop{Delta wr}	& $\Delta r/r$
			& \nop{\}}
	& \nop{wr_max}		& \nop{r_max}
			& \nop{\}}
		& \nop{u}	& \nop{->}
		& \nop{u}	& \nop{+}
		& \nop{A}	& \nop{* Gaussian(r_init=}
		& \nop{rc}	& \nop{, sigma=}
		& \nop{w}	& \nop{)}
			& \multicolumn{6}{c}{Approximate Profile of}
				& Horizon
				& \multicolumn{5}{c}{Mass}
									\\
%
%
\cline{25-29}
%
%
Name
	& $\Delta \wr$		& at $r{=}20$
			& \nop{\}}
	& $\wr_\max$		& $r_\max$
			& \nop{\}}
		& \multicolumn{10}{c}{Initial Perturbation}
			& \multicolumn{6}{c}{Scalar Field Shell ($4 \pi B \rho$)}
				& Radius
				& \multicolumn{1}{c}{BH}
				& $+$	& SF
				& $=$	& Total
									\\
%
%
\hline 
%
%
\begin{tabular}{@{}l@{}}
100.pqw5	\\
200.pqw5	\\
100o10.pqw5	\\
200o10.pqw5	\\
100o30.pqw5	\\
200o30.pqw5	
\end{tabular}
	& \begin{tabular}{@{}l@{}}
	  0.01		\\
	  0.005		\\
	  0.01		\\
	  0.005		\\
	  0.01		\\
	  0.005		
	  \end{tabular}		& \begin{tabular}{@{}l@{}}
				  0.02		\\
				  0.01		\\
				  0.02		\\
				  0.01		\\
				  0.02		\\
				  0.01		
				  \end{tabular}
			& \begin{tabular}{@{}c@{}}
			  $
			  \begin{tabular}{@{}c@{}}
			  \hbox{}	\\
			  \hbox{}	
			  \end{tabular}
			  \biggr\}
			  $
						\\
			  $
			  \begin{tabular}{@{}c@{}}
			  \hbox{}	\\
			  \hbox{}	
			  \end{tabular}
			  \biggr\}
			  $
						\\
			  $
			  \begin{tabular}{@{}c@{}}
			  \hbox{}	\\
			  \hbox{}	
			  \end{tabular}
			  \biggr\}
			  $
			  \end{tabular}
	& \begin{tabular}{@{}r@{}}
	   4		\\
			\\
	  10		\\
			\\
	  30		
	  \end{tabular}		& \begin{tabular}{@{}r@{}}
				  248		\\
						\\
				  813		\\
						\\
				  2\,780	
				  \end{tabular}
			& $
			  \left.
			  \begin{tabular}{@{}c@{}}
					  \\
					  \\
					  \\
					  \\
					  \\
			  \end{tabular}
			  \right\}
			  $
		& $P$	& ${\rightarrow}$
		& $P$	& ${+}$
		& 0.02	& ${\times} \Gaussian(r_\init{=}$
		& 20	& $, \sigma{=}$
		& 5	& $)$
			& 0.0735	& ${\times} \Gaussian(r{=}$
			& 21.8	& $, \sigma{=}$
			& 3.5	& $)$
				& 1.952
				& 0.976\,
				& $+$	& 0.641
				& $=$	& \phantom{0}1.617
									\\
%
%
\begin{tabular}{@{}l@{}}
100.pqw5i	\\
200.pqw5i	\\
\end{tabular}
	& \begin{tabular}{@{}l@{}}
	  0.01		\\
	  0.005		
	  \end{tabular}		& \begin{tabular}{@{}l@{}}
				  0.02		\\
				  0.01		
				  \end{tabular}
			& $\biggr\}$
	& \phantom{0}4		& \phantom{0\,}248
			&
		& $P$	& ${\rightarrow}$
		& $P$	& ${+}$
		& 0.02	& ${\times} \Gaussian(r_\init{=}$
		& 10	& $, \sigma{=}$
		& 5	& $)$
			& 0.0205	& ${\times} \Gaussian(r{=}$
			& 12.4	& $, \sigma{=}$
			& 3.2	& $)$
				& 1.976
				& 0.988\,
				& $+$	& 0.167
				& $=$	& \phantom{0}1.155
									\\
%
%
\begin{tabular}{@{}l@{}}
100.pqw5b	\\
200.pqw5b	
\end{tabular}
	& \begin{tabular}{@{}l@{}}
	  0.01		\\
	  0.005		
	  \end{tabular}		& \begin{tabular}{@{}l@{}}
				  0.02		\\
				  0.01		
				  \end{tabular}
			& $\biggr\}$
	& \phantom{0}4		& \phantom{0\,}248
			&
		& $P$	& ${\rightarrow}$
		& $P$	& ${+}$
		& 0.05	& ${\times} \Gaussian(r_\init{=}$
		& 20	& $, \sigma{=}$
		& 5	& $)$
			& 0.0365	& ${\times} \Gaussian(r{=}$
			& 24.4	& $, \sigma{=}$
			& 3.7	& $)$
				& 1.733
				& 0.866\,
				& $+$	& 3.405
				& $=$	& \phantom{0}4.271
									\\
%
%
\begin{tabular}{@{}l@{}}
100.pqw5c	\\
200.pqw5c	
\end{tabular}
	& \begin{tabular}{@{}l@{}}
	  0.01		\\
	  0.005		
	  \end{tabular}		& \begin{tabular}{@{}l@{}}
				  0.02		\\
				  0.01		
				  \end{tabular}
			& $\biggr\}$
	& \phantom{0}4		& \phantom{0\,}248
			&
		& $P$	& ${\rightarrow}$
		& $P$	& ${+}$
		& 0.1	& ${\times} \Gaussian(r_\init{=}$
		& 20	& $, \sigma{=}$
		& 5	& $)$
			& 0.91	& ${\times} \Gaussian(r{=}$
			& 31.0	& $, \sigma{=}$
			& 4.2	& $)$
				& --\rlap{\,\tablenotemark[1]}
				& \llap{(}0.576\rlap{)}\,
				& $+$
					& 9.737
				& $=$	& 10.313
									\\
%
%
\begin{tabular}{@{}l@{}}
200.pqw1	\\
400.pqw1	
\end{tabular}
	& \begin{tabular}{@{}l@{}}
	  0.005		\\
	  0.0025	
	  \end{tabular}		& \begin{tabular}{@{}l@{}}
				  0.01		\\
				  0.00\rlap{5}	
				  \end{tabular}
			& $\biggr\}$
	& \phantom{0}4		& \phantom{0\,}248
			&
		& $P$	& ${\rightarrow}$
		& $P$	& ${+}$
		& 0.1	& ${\times} \Gaussian(r_\init{=}$
		& 20	& $, \sigma{=}$
		& 1	& $)$
			& 1.42	& ${\times} \Gaussian(r{=}$
			& 22.71	& $, \sigma{=}$
			& 0.76	& $)$
				& 1.784
				& 0.892\,
				& $+$	& 2.692
				& $=$	& \phantom{0}3.584
									\\
%
%
\begin{tabular}{@{}l@{}}
100.pw5+qw3	\\
200.pw5+qw3	
\end{tabular}
	& \begin{tabular}{@{}l@{}}
	  0.01		\\
	  0.005		
	  \end{tabular}		& \begin{tabular}{@{}l@{}}
				  0.02		\\
				  0.01		
				  \end{tabular}
			& $\biggr\}$
	& \phantom{0}4		& \phantom{0\,}248
			&
		& \llap{$\biggl[$\,}\begin{tabular}{@{}c@{}}
		  $P$		\\
		  $Q$		
		  \end{tabular}
			& \begin{tabular}{@{}c@{}}
			  ${\rightarrow}$	\\
			  ${\rightarrow}$	
			  \end{tabular}
		& \begin{tabular}{@{}c@{}}
		  $P$		\\
		  $Q$		
		  \end{tabular}
			& \begin{tabular}{@{}c@{}}
			  ${+}$		\\
			  ${+}$		
			  \end{tabular}
		& \begin{tabular}{@{}c@{}}
		  0.02		\\
		  0.03		
		  \end{tabular}
			& \begin{tabular}{@{}c@{}}
			  ${\times} \Gaussian(r_\init{=}$	\\
			  ${\times} \Gaussian(r_\init{=}$	
			  \end{tabular}
		& \begin{tabular}{@{}c@{}}
		  20		\\
		  20		
		  \end{tabular}
			& \begin{tabular}{@{}c@{}}
			  $, \sigma{=}$		\\
			  $, \sigma{=}$		
			  \end{tabular}
		& \begin{tabular}{@{}c@{}}
		  5		\\
		  3		
		  \end{tabular}
			& \begin{tabular}{@{}c@{}}
			  $)$		\\
			  $)$		
			  \end{tabular}
\rlap{\,$\biggr]$\raisebox{1ex}{\tablenotemark[2]}}
			& 0.225	& ${\times} \Gaussian(r{=}$
			& 22.1	& $, \sigma{=}$
			& 2.5	& $)$
				& 1.872
				& 0.936\,
				& $+$	& 1.666
				& $=$	& \phantom{0}2.602
									\\
%
%
\begin{tabular}{@{}l@{}}
100.daw5c	\\
200.daw5c	
\end{tabular}
	& \begin{tabular}{@{}l@{}}
	  0.01		\\
	  0.005		
	  \end{tabular}		& \begin{tabular}{@{}l@{}}
				  0.02		\\
				  0.01		
				  \end{tabular}
			& $\biggr\}$
	& \phantom{0}4		& \phantom{0\,}248
			&
		& $A$	& ${\rightarrow}$
		& $A$	& ${+}$
		& 0.1	& ${\times} \Gaussian(r_\init{=}$
		& 20	& $, \sigma{=}$
		& 5	& $)$
			& \multicolumn{6}{c}{none}
				& 2.043
				& 1.022\,
				&	& --
				&	& \phantom{0}1.022
									\\
%
%
\end{tabular}
\end{center}
%
%
\tablenotetext[1]{
		 As discussed in the text in
		 section~\ref{sect-sample-results/other-perturbations},
		 there's no apparent horizon within this slice's numerical
		 grid; the ``black hole'' mass shown (within parentheses)
		 is actually the mass within the inner grid boundary.
		 }
\tablenotetext[2]{
		 The initial perturbation is to both $P$ and $Q$
		 for this slice.
		 }
%
%
\end{table}
%
%
\end{document}